\documentclass[]{iopart}
\usepackage{iopams}
\usepackage{times}
\usepackage{graphicx}
\usepackage{cite}
\usepackage{ae} 
\usepackage[letterpaper,bookmarks,bookmarksopen,%
            colorlinks=true,pdfstartview=Fit,urlcolor=blue]{hyperref}
\hypersetup{%
  pdftitle  = {Superconductivity of overdoped cuprates},
  pdfauthor = {PDF by E. Penev}
}

\newcommand{\openone}{\leavevmode\hbox{\small1\kern-3.3pt\normalsize1}}
\newcommand{\eps}{\epsilon}
\newcommand{\ep}{\ensuremath{\varepsilon_{\mathrm{p}}}}
\newcommand{\ed}{\ensuremath{\varepsilon_{\mathrm{d}}}}
\renewcommand{\es}{\ensuremath{\varepsilon_{\mathrm{s}}}}
\newcommand{\cuo}{\ensuremath{\textrm{CuO}_2}}
\newcommand{\cud}{\ensuremath{\textrm{Cu~3d}_{x^2-y^2}}}
\newcommand{\cus}{\textrm{Cu~4s}}
\newcommand{\tsp}{\ensuremath{t_{\mathrm{sp}}}}
\newcommand{\tpd}{\ensuremath{t_{\mathrm{pd}}}}
\newcommand{\tpp}{\ensuremath{t_{\mathrm{pp}}}}
\newcommand{\Jsd}{\ensuremath{J_{\mathrm{sd}}}}
\newcommand{\tc}{\ensuremath{T_{\mathrm{c}}}}
\newcommand{\kb}{\ensuremath{k_{_{\mathrm{B}}}}}
\newcommand{\nn}{\nonumber}
\newcommand{\Rref}[1]{reference~\cite{#1}}
\newcommand{\doi}[1]{http://dx.doi.org/#1}
\newcommand{\cm}[1]{\href{http://arXiv.org/abs/cond-mat/#1}{cond-mat/#1}}

\eqnobysec

\begin{document}

\title[Superconductivity of overdoped cuprates]{%
 Superconductivity of overdoped cuprates:
 the modern face of the ancestral two-electron exchange
}

\author{%
 T~M~Mishonov\dag\ddag,
 J~O~Indekeu\ddag\ and
 E~S~Penev\dag\
}
\address{\dag\
 \href{http://theory.phys.uni-sofia.bg}{Department of Theoretical Physics}, %
 \href{http://www.phys.uni-sofia.bg}{Faculty of Physics},\\ 
 \href{http://www.uni-sofia.bg}{Sofia University `St. Kliment Ohridski'},\\ 
 5~J~Bourchier Boulevard, BG-1164 Sofia, Bulgaria
}
\address{\ddag\
 Laboratorium voor Vaste-Stoffysica en Magnetisme, Katholieke
 Universiteit Leuven, Celestijnenlaan 200 D, B-3001 Leuven, Belgium
}

\ead{todor.mishonov@phys.uni-sofia.bg}

\begin{abstract}
The single-site two-electron exchange amplitude \Jsd\ between the Cu~4s and
Cu~3d$_{x^2-y^2}$ states is found to be the pairing mechanism of high-\tc\
overdoped cuprates.  The noninteracting part of the Hamiltonian spans the
copper Cu~4s, Cu~3d$_{x^2-y^2}$ and oxygen O~2p$_x$ and O~2p$_y$
states. Within the standard BCS treatment an explicit expression for the
momentum dependence of the gap $\Delta_{\bi{p}}$ is derived and shown to fit
the angle-resolved photoemission spectroscopy (ARPES) data.  The basic
thermodynamic and electrodynamic properties of the model [specific heat
$C(T),$ London penetration depth $\lambda(T)$] are analytically derived. These
are directly applicable to cuprates without complicating structural
accessories (chains, double \cuo\ planes, etc.).  We advocate that the pairing
mechanism of overdoped and underdoped cuprates is the same, as \tc\ displays
smooth doping dependence.  Thus, a long-standing puzzle in physics is possibly
solved.
\end{abstract}



\section{Introduction}
\label{sec:intro}

The discovery of high-temperature superconductivity~\cite{Bednorz:86} in
cuprates and the subsequent ``research rush'' have led to the appearance of
about 100~000 papers to date~\cite{Ginzburg:00}. Virtually every fundamental
process known in condensed matter physics was probed as a possible mechanism
of this phenomenon. Nevertheless, none of the theoretical efforts resulted in
a coherent picture~\cite{Ginzburg:00}. For the conventional superconductors
the mechanism was known to be the interaction between electrons and
crystal-lattice vibrations, but the development of its theory lagged behind
the experimental findings.  The case of cuprate high-\tc\ superconductivity
appears to be the opposite: we do not convincingly know which mechanism is to
be incorporated in the traditional Bardeen-Cooper-Schrieffer (BCS)
theory~\cite{BCS}.  Thus the path to high-\tc\ superconductivity in cuprates,
perhaps carefully hidden or well-forgotten, has turned into one of the
long-standing mysteries in physical science.

Features of the electronic spectrum of the \cuo\ plane, \fref{fig:1}(a), the
structural detail responsible for the superconductivity of the cuprates, have
become accessible from the angle-resolved photoemission spectroscopy
(ARPES)~\cite{Aebi:01,Campuzano:02}. Thus, any theory which pretends to
explain the cuprate superconductivity is bound to include these features and
account for them consistently. A number of extensive reviews over the past
years have been devoted to that theoretical
problem~\cite{Scalapino:95,Schrieffer:95,Plakida:95,Annet:96,Koltenbah:96,Ruvalds:96,Markiewicz:97,Brusov:99,Wilson:00,Szotek:01,Chubukov:02,Carlson:02}.
For further related discussion we also refer the reader to the
review~\cite{Rigamonti:98} on NMR-NQR spectroscopies in high-\tc\
superconductors.

In contrast with all previous proposals, we have advanced in \Rref{highway}
the \emph{intra-atomic} exchange~\cite{ref:s-d} of two electrons between the
4s and 3d$_{x^2-y^2}$ states of the Cu atom as the origin of high-\tc\
superconductivity in the layered cuprates and have shown that the basic
spectroscopic and thermodynamic experiments can be explained by it. Previously
only \emph{inter-atomic} Heitler-London-type~\cite{HeitlerLondon} two-electron
exchange~\cite{Mishonov:97,Mishonov:98,Mishonov:02a} has been discussed. Thus,
the present work is the unabridged version of our theory announced in
\Rref{highway}.  It builds upon the standard
Bloch-H\"uckel~\cite{Bloch,Hueckel,Slater,Labbe:87} (tight-binding)
approximation to the electronic band structure of the \cuo\ plane, developed
in an earlier work~\cite{Mishonov:00}.  We derive an analytical expression for
the BCS kernel, or pairing potential $V_{\bi{p}\bi{p}'}.$ For the case of the
s-d pairing the analytical solution is compared to the ARPES data.  Extensive
discussion is also provided to help the juxtaposition of our theory with other
models. Finally, exact expressions within the s-d model are given for the
specific heat, London penetration depth, Cooper-pair effective mass and Hall
constant of the vortex-free Meissner-Ochsenfeld phase.

\section{Lattice Hamiltonian}
\label{sec:model}

The electronic properties of materials are strongly influenced by the local
environment and in this sense the electronic features are local physics. The
simplest possible model for high-\tc\ superconductivity contains
single-particle and correlated two-electron hoppings between nearest
neighbours and next-nearest neighbours. Formally, this is an expansion of the
many-particle Hamiltonian containing two- and four-fermion operators. The
two-fermion Hamiltonian determines the band structure, briefly considered in
subsection~\ref{sec:tfbmn}, while the four-fermion terms
(subsection~\ref{sec:HL}) determine the pairing interaction, and lead to the
gap equations considered in section~\ref{sec:RH}.

\subsection{The four-band model in a nutshell}
\label{sec:tfbmn}

\begin{figure}[t]
\centering\includegraphics[width=0.7\columnwidth]{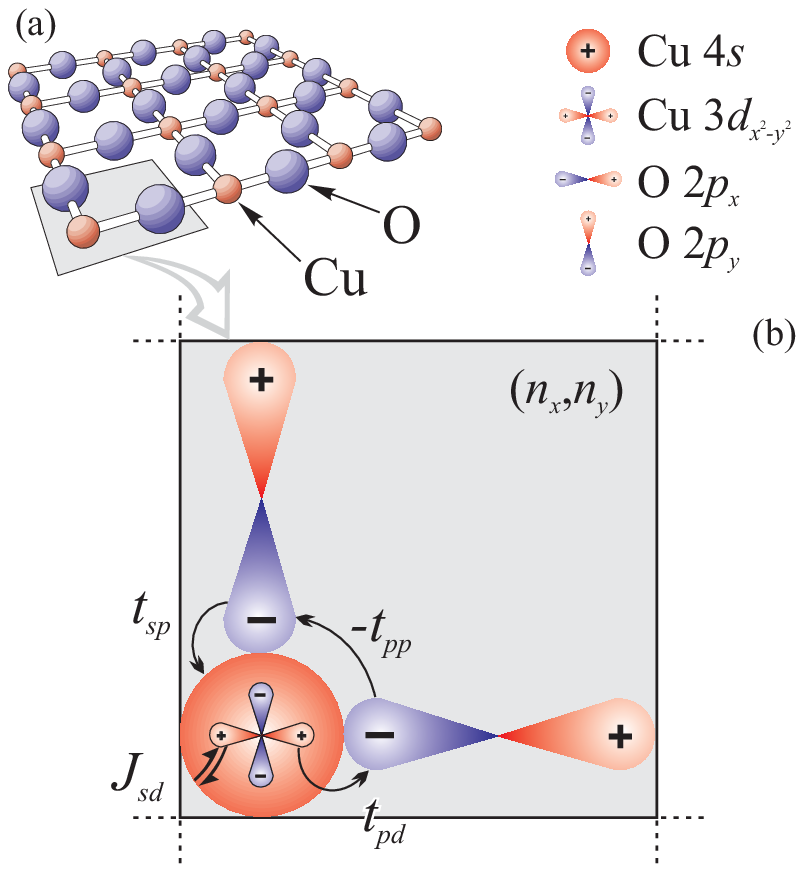}
\caption{(a) Ball-and-stick model of the \cuo\ plane. The shaded
square is the unit cell indexed by $\bi{n}=(n_x,n_y),\,
n_{x,y}=0,\pm1,\pm2,\dots$. (b) The LCAO basis set: A single
electron hops from the 3d atomic orbital to 2p$_x$ with amplitude
\tpd, contained in $\hat{\cal H}_{\mathrm{BH}}$. From 2p$_x$ to
2p$_y$ the hopping amplitude is \tpp, and from there to 4s the
hopping amplitude is \tsp. Correlated hopping of two electrons in
opposite directions between 3d and 4s with amplitude \Jsd\ is
depicted as a double arrow (see the discussion in
sections~\protect\ref{sec:sd} and \protect\ref{sec:signJ}).
\label{fig:1} }
\end{figure}

Every high-\tc\ superconductor has its specific properties. It is strongly
believed, however, that the main features of the electronic band structure of
the \cuo\ plane are adequately described by the four-band model spanning the
\cud, \cus, O~2p$_x$ and O~2p$_y$ orbitals, \fref{fig:1}(b). In the spirit of
the Bloch-H\"uckel (BH) model, using Jordan's second quantization language, we
introduce Fermi annihilation operators for an electron with spin projection
$\alpha$ at a particular orbital, respectively,
$\hat{D}_{\bi{n}\alpha},
 \hat{S}_{\bi{n}\alpha},
 \hat{X}_{\bi{n}\alpha},$ and $ \hat{Y}_{\bi{n}\alpha}$
in the unit cell with index $\bi{n}=(n_x,n_y).$ It is convenient
to introduce a multicomponent Fermi creation operator in momentum
space,
$\hat{\Psi}_{\bi{p}\alpha}^{\dag}= (\hat{D}_{\bi{p}\alpha}^{\dag},
 \hat{S}_{\bi{p}\alpha}^{\dag},
 \hat{X}_{\bi{p}\alpha}^{\dag},
 \hat{Y}_{\bi{p}\alpha}^{\dag})$.
In this notation the one-electron BH Hamiltonian reads
\begin{equation}
 \hat{\cal{H}}_{\rm BH}'= \hat{\cal{H}}_{\rm BH} - \mu\hat{\cal{N}} =
 \sum_{\bi{p},\alpha}\hat{\Psi}_{\bi{p}\alpha}^{\dag}
  (H_{\rm BH}-\mu\openone_{4\times 4})\hat{\Psi}_{\bi{p}\alpha},
\label{eq:BH}
\end{equation}
where $\mu$ is the chemical potential, and (cf.
\Rref{Mishonov:00})
\begin{equation}
H_{\rm BH}
    = \left(
        \begin{array}{cccc}
            \eps_{\mathrm{d}}   & 0        & \tpd s_x      & -\tpd s_y     \\
                 0   & \eps_{\mathrm{s}}   & \tsp s_x      & \tsp s_y      \\
            \tpd s_x & \tsp s_x & \eps_{\mathrm{p}}        & -\tpp s_x s_y \\
            -\tpd s_y& \tsp s_y & -\tpp s_x s_y & \eps_{\mathrm{p}}
        \end{array}
    \right);
\label{eq:4x4}
\end{equation}
$\eps_{\mathrm{d}},$ $\eps_{\mathrm{s}},$ and $\eps_{\mathrm{p}}$
are the single-site energies of the \cud, \cus, O~2p$_x$ and
O~2p$_y$ states, respectively. The hopping amplitudes between
these states are \tsp, \tpd\ and \tpp, \fref{fig:1}(b). Note, that
because of the orbital orthogonality $t_{\mathrm{sd}}=0$. For
brevity, we have introduced also the notation
\begin{eqnarray}
s_x  &= 2\sin(p_x/2), \quad  s_y =2\sin(p_y/2), \quad \bi{s}=(s_x,s_y),\nn\\
x    &= \sin^2(p_x/2),\quad  y   =\sin^2(p_y/2).
\label{eq:s}
\end{eqnarray}

From a classical point of view, the \cud\ state corresponds to a
circular electron rotation in the \cuo\ plane while the \cus\
state corresponds to a classical ensemble of electrons of zero
angular momentum continuously falling to the nucleus. Pictorially,
the s-electrons fall to the nuclei like comets, but after the
impact the turning point of their motion is very far from the
nucleus. This is the reason why \tsp\ is considerably larger than
\tpd. The transfer amplitude \tpp\ is the smallest one since the
hopping to the next-nearest neighbour requires a tunnelling
through free space. As a rule, the electron band calculations
significantly overestimate \tpp, but the latter can be reliably
calculated using the surface integral method, cf.
\Rref{Mishonov:00}. Even for the largest transfer integrals \tsp\
and \tpd, which determine the bandwidth of the conduction band,
the \textit{ab initio} calculations give a factor 2 or even 3
``overbinding''. Nonetheless, the band calculations substantiate
this choice for the LCAO (linear combination of atomic orbitals)
basis set and provide an adequate language for interpretation. In
the end, these parameters should be determined by fitting to the
spectroscopy data and be treated in the model lattice Hamiltonian
as phenomenological parameters of the microscopic many-body
theory. We shall briefly recall some basic properties of the
four-band model as derived in \Rref{Mishonov:00}.

Let $\eps_{b,\bi{p}}$ and $\Psi_{b,\bi{p}}$ be the eigenvalues and
the corresponding eigenvectors of the BH Hamiltonian, $H_{\rm
BH}\Psi_{b,\bi{p}} = \eps_{b,\bi{p}}\Psi_{b,\bi{p}}$, where
$b=1,\ldots\!, 4$ is the band index. For the ``standard model'',
$\eps_{\mathrm{p}}<\eps_{\mathrm{d}}<\eps_{\mathrm{s}},$ the
lowest energy band, $b=1,$ is an oxygen bonding band having a
minimum at the $(\pi,\pi)$ point. The next band, $b=2,$ is a
narrow ``nonbonding'' oxygen band with an exactly (within the
framework of the model) zero dispersion along the
$(0,0)$-$(\pi,0)$ direction, i.e., this band is characterized by
an extended Van Hove singularity. The conduction band, $b=3$,
is a nearly half-filled \cud\ band with the analytical eigenvector
\begin{equation}
\fl \tilde\Psi_{3,\bi{p}}= \left(\!\!
    \begin{array}{c}
        D_{3,\bi{p}} \\
        S_{3,\bi{p}} \\
        X_{3,\bi{p}} \\
        Y_{3,\bi{p}}
    \end{array}
\!\!\right) = \left(\!\!
    \begin{array}{c}
        -\es\,\ep^2 + 4 \ep\,\tsp^2\,(x+y) - 32 \tpp\,\tau^2_{sp}\, xy\\
        -4\ep\,\tsp\,\tpd\,(x-y) \\
        -(\es\,\ep - 8\tau^2_{\mathrm{sp}}\,y)\,\tpd\, s_x \\
         (\es\,\ep - 8\tau^2_{\mathrm{sp}}\,x)\,\tpd\, s_y
    \end{array}
\!\!\right), \label{eq:vector}
\end{equation}
where the $\varepsilon$'s denote the energies measured relative to
their respective atomic levels: $\es = \eps-\eps_{\mathrm{s}},$ $\ep =
\eps-\eps_{\mathrm{p}}$, $\ed = \eps-\eps_{\mathrm{d}},$ and
$\tau_{\mathrm{sp}}^2 =\tsp^2 -\es\,\tpp/2$.  The topmost band, $b=4$,
is an empty \cus\ band. In elemental metals like Cu and Fe, the 4s
band is a wide conduction band, but for the \cuo\ plane it is
completely ``oxidized''. Having the analytical eigenvector we can
calculate the corresponding eigenvalue:
\begin{equation}
\eps_{3,\bi{p}}=\frac{\langle\tilde\Psi_{3,\bi{p}}\,\vert H_{\rm
BH}\vert\tilde\Psi_{3,\bi{p}}\rangle}
{\langle\tilde\Psi_{3,\bi{p}}\,\vert\tilde\Psi_{3,\bi{p}}\rangle}.
\label{eq:Rayleigh}
\end{equation}
If necessary, the nonorthogonality of the atomic orbitals at neighbouring
atoms can be easily taken into account. In this case the normalizing
denominator in the above equation reads (for arbitrary band index)
\begin{eqnarray}
\langle\tilde\Psi_{\bi{p}}\,\vert\tilde\Psi_{\bi{p}}\rangle  = &
  D^2_{\bi{p}} + S^2_{\bi{p}} + X^2_{\bi{p}} + Y^2_{\bi{p}}
  + 2 g_{\mathrm{pd}} s_x D_{\bi{p}} X_{\bi{p}}
  - 2 g_{\mathrm{pd}} s_y D_{\bi{p}} Y_{\bi{p}} \nn\\
& + 2 g_{\mathrm{sp}} s_x S_{\bi{p}} X_{\bi{p}}
  + 2 g_{\mathrm{sp}} s_y S_{\bi{p}} Y_{\bi{p}}
  - 2 g_{\mathrm{pp}} s_x s_y X_{\bi{p}} Y_{\bi{p}},
\label{eq:norm2}
\end{eqnarray}
where the ``metric tensor'' $g_{ij}$ is given by the integral
\begin{equation}
g_{ij} = \int \psi^*_i(\bi{r})\psi_j(\bi{r}-\bi{R}) \rmd\bi{r},
\label{eq:metrix}
\end{equation}
where $\psi^*_i(\bi{r})$ and $\psi_j(\bi{r}-\bi{R})$ are the atomic wave
functions, and $\bi{R}$ is the inter-atomic distance.  The phases are chosen
such that all overlap integrals $g_{\mathrm{pd}},$ $g_{\mathrm{sp}},$ and
$g_{\mathrm{pp}}$ be positive parameters, like the hopping integrals
$t_{\mathrm{pd}},$ $t_{\mathrm{sp}},$ and $t_{\mathrm{pp}}.$ Note that these
provisions apply only to the single-particle spectrum. As long as one deals
with a single conduction band, all Bloch states are orthogonal and the further
treatment of the second-quantized Hamiltonian proceeds in the standard way.

Thus, using the Rayleigh quotient iteration for
equations~\eref{eq:4x4}--\eref{eq:Rayleigh} one can obtain
numerically the eigenvalue and the eigenvector. The band energies
$\eps\equiv\eps_{b,\bi{p}}$ satisfy the secular equation
\begin{equation}
 \det \left(H_{\mathrm{BH}}-\eps\openone_{4\times4}\right) =
 {\cal A} xy + {\cal B} (x+y)+{\cal C}=0,
\label{eq:cec}
\end{equation}
where the energy-dependent coefficients read~\cite{Mishonov:00}:
\begin{eqnarray}
 {\cal A}(\eps) &=  16\left(4\tpd^2\tsp^2+2\tsp^2\tpp\ed-2\tpd^2\tpp\es
                    -\tpp^2\ed\es\right),                 \nn\\
 {\cal B}(\eps) &=  -4\ep\left(\tsp^2\ed+\tpd^2\es\right), \label{eq:ABC}\\
 {\cal C}(\eps) &= \ed\es\ep^2. \nn
\end{eqnarray}
Furthermore we introduce the normalized eigenvector $\Psi_{
b,\bi{p}} = \tilde\Psi_{b,\bi{p}}/\|\tilde\Psi_{b,\bi{p}} \|$ and
write the noninteracting Hamiltonian in diagonal form,
\begin{equation}
 \hat{\cal{H}}_{\rm BH}'=  \sum_{b,\bi{p},\alpha}(\eps_{b,\bi{p}}-\mu)
 \hat{c}_{b,\bi{p}\alpha}^{\dag}\hat{c}_{b,\bi{p}\alpha}.
\label{eq:diagonal}
\end{equation}
The Fermi operators in real-space representation can be easily
expressed using the band representation,
\begin{equation}
 \hat{\Psi}_{\bi{n}\alpha} \equiv \left(\!\!
    \begin{array}{c}
        \hat{D}_{\bi{n}\alpha} \\
        \hat{S}_{\bi{n}\alpha} \\
        \hat{X}_{\bi{n}\alpha} \\
        \hat{Y}_{\bi{n}\alpha}
    \end{array}
 \!\!\right)
    = \frac{1}{\sqrt{N}} \sum_{b,\bi{p}} \e^{\rmi\bi{p}\cdot\bi{n}}
 \left(\!\!
    \begin{array}{c}
        D_{b,\bi{p}} \\
        S_{b,\bi{p}} \\
        \rme^{\rmi\varphi_x} X_{b,\bi{p}}  \\
        \rme^{\rmi\varphi_y} Y_{b,\bi{p}}
    \end{array}
 \!\!\right)\hat{c}_{b,\bi{p}\alpha},
\label{eq:space}
\end{equation}
where $N$ is the number of unit cells, and the two phases in the
right-hand side of the equation read $\varphi_x =
\frac{1}{2}(p_x-\pi)$ and $\varphi_y = \frac{1}{2}(p_y-\pi).$ This
transformation will be used in the next subsection for deriving
the interaction Hamiltonian.

\subsection{The Heitler-London \& Schubin-Wonsowsky-Zener interaction}
\label{sec:HL}

The Heitler-London (HL) interaction Hamiltonian describes the
(intra- and inter-atomic) two-electron exchange. It comprises
four parts~\cite{Mishonov:97,Mishonov:98} corresponding to
$\cus\leftrightarrow\textrm{O~2p}\sigma$,
$\textrm{O~2p}\sigma\leftrightarrow\cud$,
$\textrm{O~2p}_x\leftrightarrow\textrm{O~2p}_y$, and
$\cud\leftrightarrow\cus$ exchanges with transition amplitudes
$J_{\mathrm{sp}}$, $J_{\mathrm{pd}}$, $J_{\mathrm{pp}},$ and \Jsd,
respectively:
\begin{eqnarray}
 \fl  \hat{\cal{H}}_{\rm HL} = -\Jsd \sum_{\bi{n},\alpha\beta}
    \hat{S}^{\dag}_{\bi{n}\alpha}\hat{D}^{\dag}_{\bi{n}\beta}
     \hat{S}_{\bi{n}\beta}\hat{D}_{\bi{n}\alpha} \nn \\
 \fl \qquad\quad
 - J_{\mathrm{sp}} \sum_{\bi{n},\alpha\beta}
   \Big[
    \hat{S}^{\dag}_{\bi{n}\alpha}\hat{X}^{\dag}_{\bi{n}\beta}
     \hat{S}_{\bi{n}\beta}\hat{X}_{\bi{n}\alpha}
      +\hat{S}^{\dag}_{\bi{n}\alpha}\hat{Y}^{\dag}_{\bi{n}\beta}
       \hat{S}_{\bi{n}\beta}\hat{Y}_{\bi{n}\alpha} \nn \\
 \fl  \qquad \qquad \quad
  +\hat{S}^{\dag}_{(n_x+1,n_y)\alpha} \hat{X}^{\dag}_{\bi{n}\beta}
     \hat{S}_{(n_x+1,n_y)\beta} \hat{X}_{\bi{n}\alpha}
    +\hat{S}^{\dag}_{(n_x,n_y+1)\alpha} \hat{Y}^{\dag}_{\bi{n}\beta}
     \hat{S}_{(n_x,n_y+1)\beta}\hat{Y}_{\bi{n}\alpha}
  \Big] \nn\\
 \fl \qquad\quad
  - J_{\mathrm{pd}} \sum_{\bi{n},\alpha\beta}
  \Big[
   \hat{D}^{\dag}_{\bi{n}\alpha} \hat{X}^{\dag}_{\bi{n}\beta}
   \hat{D}_{\bi{n}\beta}\hat{X}_{\bi{n}\alpha}
  +\hat{D}^{\dag}_{\bi{n}\alpha}\hat{Y}^{\dag}_{\bi{n}\beta}
   \hat{D}_{\bi{n}\beta} \hat{Y}_{\bi{n}\alpha} \nn  \\
 \fl  \qquad \qquad \quad
  +\hat{D}^{\dag}_{(n_x+1,n_y)\alpha}\hat{X}^{\dag}_{\bi{n}\beta}
     \hat{D}_{(n_x+1,n_y)\beta}\hat{X}_{\bi{n}\alpha}
    +\hat{D}^{\dag}_{(n_x,n_y+1)\alpha}\hat{Y}^{\dag}_{\bi{n}\beta}
     \hat{D}_{(n_x,n_y+1)\beta}\hat{Y}_{\bi{n}\alpha}
  \Big] \nn\\
 \fl  \qquad\quad
  -J_{\mathrm{pp}} \sum_{\bi{n},\alpha\beta}
  \Big[
   \hat{X}^{\dag}_{\bi{n}\alpha}\hat{Y}^{\dag}_{\bi{n}\beta}
   \hat{X}_{\bi{n}\beta} \hat{Y}_{\bi{n}\alpha}
  +\hat{X}^{\dag}_{\bi{n}\alpha} \hat{Y}^{\dag}_{(n_x+1,n_y)\beta}
   \hat{X}_{\bi{n}\beta} \hat{Y}^{}_{(n_x+1,n_y)\alpha}   \nn  \\
 \fl  \qquad \qquad \quad
   +\hat{X}^{\dag}_{(n_x,n_y+1)\alpha} \hat{Y}^{\dag}_{\bi{n}\beta}
   \hat{X}_{(n_x,n_y+1)\beta} \hat{Y}_{\bi{n}\alpha}  \nn \\
 \fl  \qquad \qquad \quad
   +\hat{X}^{\dag}_{(n_x,n_y+1)\alpha}\hat{Y}^{\dag}_{(n_x+1,n_y)\beta}
   \hat{X}_{(n_x,n_y+1)\beta}\hat{Y}_{(n_x+1,n_y)\alpha}
  \Big].
\label{eq:interaction}
\end{eqnarray}
Let us now analyze the structure of the total electron Hamiltonian
$\hat{\cal{H}}' = \hat{\cal{H}}_{\rm BH}' +\hat{\cal{H}}_{\rm HL}$. In terms
of the Fermi operators $\hat{\Psi}_{i\alpha},$ corresponding to the atomic
orbitals, $\hat{\cal{H}}'$ reads:
\begin{eqnarray}
\hat{\cal{H}}' =& \sum_{i,\alpha}(\eps_i
 -\mu)\hat{\Psi}_{i\alpha}^{\dag}\hat{\Psi}_{i\alpha}
 -\!\sum_{i<j,\alpha}
\left(\tilde t_{ji}\hat{\Psi}_{j\alpha}^{\dag}\hat{\Psi}_{i\alpha}
 + \tilde t_{ji}^*\hat{\Psi}_{i\alpha}^{\dag}\hat{\Psi}_{j\alpha}\right)
 \nn \\
 &-\sum_{i<j,\alpha\beta}J_{ij}
 \hat{\Psi}_{i\beta}^{\dag}\hat{\Psi}_{j\alpha}^{\dag}
 \hat{\Psi}_{i\alpha} \hat{\Psi}_{j\beta},
\label{eq:general}
\end{eqnarray}
where $\tilde t_{ji}= t_{ji}\textrm{e}^{\rmi \phi_{ji}}$, $t_{ji}=t_{ij}$ and
$\phi_{ji}=\phi_j-\phi_i$ is the phase difference between the $i$th and $j$th
atomic orbitals in the overlapping domain. Roughly speaking, onto every
single-electron hopping amplitude $t_{ij}$ one can map a corresponding
two-electron hopping amplitude $J_{ij}$. The case of a strong electron
correlation implies that $J_{ij}$ could be of the order of $t_{ij}$. Thus, one
can expect that the following inequalities hold true
$J_{\mathrm{pp}}<J_{\mathrm{pd}}<J_{\mathrm{sp}}<\Jsd.$

In fact, the s-d exchange is the basic process responsible for the magnetism
of transition metals; see for example \Rref{ref:s-d}. It was understood since
the dawn of quantum physics that the mechanism of
ferromagnetism~\cite{Heisenberg} is the two-electron exchange owing to the
electron correlations~\cite{ref:sd-first}.

Here we shall add a few words in retrospect concerning the two-electron
correlation parameterized by $J_{ij}$ in \eref{eq:general}. Probably the first
two-electron problem was Bohr's consideration of the He atom~\cite{Bohr}
(cf. references \cite{Langmuir:21,vanVleck:22}) in which two electrons have
opposite coordinates $\bi{r}_2=-\bi{r}_1$ and momenta
$\bi{p}_2=-\bi{p}_1$. For a purely radial motion, such a fall to the nucleus
is stable and many years after Bohr's prediction double Rydberg states, with
an effective $\mathrm{Ry}_\mathrm{eff}=(2-1/4)\,\mathrm{Ry},$ were discovered
by electron energy loss spectroscopy \cite{Read:82}. These double Rydberg states with
opposite electron momenta can be considered as proto-forms of the Cooper
pairs. Interestingly, in 1914, Sir~J~J~Thomson proposed~\cite{Thompson:14}
(cf. also the textbook~\cite{Blakemore:85}) that electric charge can propagate
as electron doublets---another proto-form of the local (Ogg-Schafroth)
pairs~\cite{Ogg:46,Schafroth:55}. Before the appearance of quantum mechanics,
Lewis~\cite{Lewis:16} and Langmuir~\cite{Langmuir:19} introduced the idea of
electron doublets in order to explain the nature of the chemical bond.  Nearly
at the same time Parson~\cite{Parson:15} came to the conclusion that ``an
electron is not merely an electron charge but a small magnet'' or in his
terminology ``a magneton'', cf.~\Rref{Lewis:16}.  Later, in 1926, Lewis
introduced also the notion of a photon~\cite{Lewis:26} without any reliable
theoretical background at the time.

In the era of new quantum mechanics Heitler and London~\cite{HeitlerLondon}
realized the idea of electron doublets~\cite{Wilson:84} and convincingly
demonstrated how the two-particle correlation owing to a strong Coulomb
repulsion can lead to a decrease of the energy, and by virtue of the
Hellmann-Feynman theorem, to inter-atomic attraction for the singlet state of
the electron doublet. The original Heitler-London calculation, which is
nowadays interpreted in every textbook in quantum mechanics and/or quantum
chemistry, gives indeed a wrong sign of the exchange energy for very large
inter-atomic distances but, in principle, there are no conceptual difficulties
in the Heitler-London theory. The exchange energy $J$ was
represented~\cite{HerringFlicker} as a surface integral in the two-electron
six-dimensional space $(\bi{r}_1,\,\bi{r}_2)$ and this was shown to be an
asymptotically exact result, cf. also reference~\cite{Herring:66}.  The
surface integral method gives amazingly accurate results (cf. the excellent
monograph by Patil and Tang~\cite{PatilTang}) even if the exchange energy is
of the order of the energies typical for solid state phenomena.
Unfortunately, this method, that ought to be applied to \textit{ab initio}
calculated (e.g., from density functional theory (DFT)~\cite{DFT}) wave
functions, is barely known in the solid state community (although a very
recent work by Gor'kov and Krotkov~\cite{Gorkov:02} indicates that it is not
completely forgotten).

This is one of the reasons why the $t$ and $J$ transfer integrals have been
treated phenomenologically just as fitting parameters of the theory. A
valuable discussion on a similar scope of ideas has recently been given by
Brovetto, Maxia and Salis~\cite{Brovetto:00} but it may well not be the only
case. In order to ease comparison of the HL Hamiltonian with the other types
discussed in the search of a theory of high-\tc\ superconductivity we shall
rewrite it in terms of spin variables.

The grounds for our theory have been set first by Schubin and Wonsowsky and
later in more clear notions and notation by Zener~\cite{ref:s-d}. The s-d
two-electron exchange is the intra-atomic version of the HL interaction. Both
of those 4-fermion interactions due to Heitler-London \&
Schubin-Wonsowsky-Zener can in principle mediate superconductivity and
magnetism.

\subsubsection{Spin variables}

Let us introduce the spin operator $\hat\bi{S}_i$ and particle number operator
$\hat n_i$ for each atomic orbital,
\begin{equation}
 \hat\bi{S}_i  = \hat \Psi_i^{\dag}\,\frac{\bsigma}{2}\,\hat\Psi_i,\qquad
 \hat n_i          = \hat\Psi_i^\dag\, \sigma_0\, \hat\Psi_i,\qquad
\hat \Psi_i^{\dag} = \left(\hat\Psi_{i\uparrow}^\dag,
                     \hat \Psi_{i\downarrow}^\dag\right),
\end{equation}
where $\sigma_0=\openone_{2\times2}$ and $\bsigma$ are the Pauli sigma
matrices, and the first two formulae imply summation over the spin
indices. Introducing also the spin exchange operator $\hat P_{ij},$
\begin{equation}
 P\hat\Psi_{i\alpha}\hat\Psi_{j\beta}
       =\hat\Psi_{i\beta}\hat\Psi_{j\alpha},\qquad
 \hat P_{ij}=
 \sum_{\alpha\beta}(\hat\Psi_{i\alpha}\hat\Psi_{j\beta})^\dag
  P\,\hat\Psi_{i\alpha}\hat\Psi_{j\beta},
\label{eq:P}
\end{equation}
we can rewrite the HL Hamiltonian \emph{per bond}
as~\cite{Dirac,Feynman,Schwinger}
\begin{equation}
-J\sum_{\alpha\beta}
 \hat{\Psi}_{i\beta}^{\dag}\hat{\Psi}_{j\alpha}^{\dag}
 \hat{\Psi}_{i\alpha} \hat{\Psi}_{j\beta}=
 J\hat P_{ij}=2J\left(\hat \bi{S}_i\cdot\hat \bi{S}_j
   +\frac{1}{4}\hat n_i\hat n_j\right).
\label{eq:bond}
\end{equation}
We should stress that in the $t$-$J$ model the term $\propto \hat n_i\hat n_j$
enters with negative sign~\cite{Harris:67,SpalekHonig}. Let us also provide
the ``mixed'' representation:
\begin{eqnarray}
 2\hat\bi{S}_i\cdot\hat\bi{S}_j
  &= \hat S_{i,x}\left(\hat{\Psi}_{j\uparrow}^{\dag}\hat{\Psi}_{j\downarrow}
     +\hat{\Psi}_{j\downarrow}^{\dag}\hat{\Psi}_{j\uparrow}\right) \nn \\
&\quad+\hat
S_{i,y}\left(-i\hat{\Psi}_{j\uparrow}^{\dag}\hat{\Psi}_{j\downarrow}
     +i\hat{\Psi}_{j\downarrow}^{\dag}\hat{\Psi}_{j\uparrow}\right)
     +\hat S_{i,z}\left(\hat n_{j\uparrow}-\hat n_{j\downarrow}\right) \nn\\
  &= \hat S_{i,+} \hat{\Psi}_{j\downarrow}^{\dag}\hat{\Psi}_{j\uparrow}
     + \hat S_{i,-} \hat{\Psi}_{j\uparrow}^{\dag}\hat{\Psi}_{j\downarrow}
     +\hat S_{i,z}\left(\hat n_{j\uparrow}-\hat n_{j\downarrow}\right),
\label{eq:mixed}
\end{eqnarray}
where $\hat
n_{j\uparrow}\equiv\hat{\Psi}_{j\uparrow}^{\dag}\hat{\Psi}_{j\uparrow},$ and
$\hat S_{i,+} = \hat{\Psi}_{i\uparrow}^{\dag}\hat{\Psi}_{i\downarrow} = \hat
S_{i,-}^{\dag}$.  Note that (\ref{eq:bond}) implies a purely orbital motion
without spin flip: two electrons exchange their orbitals and only the spin
indices reflect this correlated hopping. For $J>0,$ the HL Hamiltonian has a
singlet ground state
\begin{equation}
\fl
 |\textrm{S}\rangle= \frac{1}{\sqrt{2}}
  (\hat\Psi_{i\uparrow}^{\dag}\hat\Psi_{j\downarrow}^{\dag}
 -\hat\Psi_{i\downarrow}^{\dag}\hat\Psi_{j\uparrow}^{\dag})
  |\textrm{vac}\rangle,\qquad
 \hat{\Psi}_{i\alpha}|\textrm{vac}\rangle = 0,\qquad
 \langle\textrm{vac}|\textrm{vac}\rangle = 1,
\end{equation}
with eigenvalue $-J.$ The lowering in energy of the singlet state, having a
symmetric orbital wave function, is of purely kinetic origin related to the
delocalization of the particles at different orbitals. Symbolically, the
``location'' of the (approximately) localized electron doublet(s) in the
structure signature of a molecule is designated by a colon, e.g., H:H for the
H$_2$ molecule. This Lewis notation for the valence bond with energy $-J$ (or
four-Fermion terms in the second quantization language) is an important
ingredient of the chemical intuition. In principle, such an exchange lowering
is expected to exist for Bose particles as well. For electrons, however, we
have triplet excited states
\begin{eqnarray}
 |\textrm{T+1}\rangle &= \hat\Psi_{i\uparrow}^{\dag}\hat\Psi_{j\uparrow}^{\dag}
                         |\textrm{vac}\rangle, \nn \\
 |\textrm{T\,0}\rangle &= \frac{1}{\sqrt{2}}
                (\hat\Psi_{i\uparrow}^{\dag}\hat\Psi_{j\downarrow}^{\dag}
                + \hat\Psi_{i\downarrow}^{\dag}\hat\Psi_{j\uparrow}^{\dag})
                |\textrm{vac}\rangle,  \\
 |\textrm{T\,--1}\rangle
 &=\hat\Psi_{i\downarrow}^{\dag}\hat\Psi_{j\downarrow}^{\dag}
                |\textrm{vac}\rangle,\nn
\end{eqnarray}
with eigenvalue $J$. In the present work we consider the parameter $J$
to be \emph{positive if it corresponds to antiferromagnetism, or
pairing in the singlet channel}. Thus the singlet-triplet splitting
for the single-bond HL Hamiltonian (\ref{eq:bond}) is $2J$. Similarly,
the bonding-antibonding splitting for the single-particle hopping
Hamiltonian $-t\sum_{\alpha}
(\hat{\Psi}_{j\alpha}^{\dag}\hat{\Psi}_{i\alpha}
+\hat{\Psi}_{i\alpha}^{\dag}\hat{\Psi}_{j\alpha})$ is $2t,$ and the
energy threshold for creation of a pair of normal carriers, considered
in the next section, is $2\Delta$.  Besides stemming from bare inter-
and intra-atomic processes, two-electron hopping amplitudes $J$ can be
created by strong correlations~\cite{Fulde:95} within the Hubbard
model. For a nice review on this subject the reader is referred to the
work by Spalek and Honig~\cite{SpalekHonig}.

\section{Reduced Hamiltonian and the BCS gap equation}
\label{sec:RH}

Substituting the Fermi operators $\hat{\Psi}_{\bi{n}\alpha},$
equation~(\ref{eq:space}), into equation~(\ref{eq:interaction})
one obtains the HL interaction Hamiltonian in a diagonal band
representation. For the case of zero electric
current~\cite{Ketterson:99}, solely the reduced Hamiltonian
$\hat{\cal{H}}_{\mbox{\scriptsize HL-R}},$ including creation and
annihilation operators with opposite momenta only, has to be taken
into account:
\begin{equation}
\hat{\cal{H}}_{\mbox{\scriptsize HL-R}}  =
 \frac{1}{2N} \sum_{b,\bi{p}} \sum_{b'\!,\bi{p}'}\sum_{\alpha\beta}
 \hat{c}_{b,\bi{p}\beta}^{\dag}\hat{c}_{b,-\bi{p}\alpha}^{\dag}\,
  V_{b,\bi{p};b'\!,\bi{p}'}\,
 \hat{c}_{b'\!,-\bi{p}'\alpha}\hat{c}_{b'\!,\bi{p}'\beta}.
\end{equation}
For singlet superconductors it is necessary to take into account
the pairing with opposite spins, thereby the \emph{total reduced}
Hamiltonian reads
\begin{equation}
\fl
 \hat{\cal{H}}_{\rm R}' =
 \sum_{b,\bi{p},\alpha}\eta_{b,\bi{p}}\,
  \hat{c}_{b,\bi{p}\alpha}^{\dag}\hat{c}_{b,\bi{p}\alpha}
 +\frac{1}{N}\sum_{b,\bi{p}} \sum_{b'\!,\bi{p}'}V_{b,\bi{p};b'\!,\bi{p}'}\,
 \hat{c}_{b,\bi{p}\uparrow}^{\dag}\hat{c}_{b,-\bi{p}\downarrow}^{\dag}
 \hat{c}_{b'\!,-\bi{p}'\downarrow}\hat{c}_{b'\!,\bi{p}'\uparrow},
\label{eq:RHtot}
\end{equation}
where $\eta_{b,\bi{p}}\equiv\eps_{b,\bi{p}}-\mu$ are the band
energies measured from the chemical potential~\cite{Ketterson:99}.
Hence the BCS equation~\cite{BCS} for the superconducting gap
takes the familiar form
\begin{equation}
\Delta_{b,\bi{p}} = \frac{1}{N}\sum_{b'\!,\bi{p}'}(-
        V_{b,\bi{p};b'\!,\bi{p}'})
        \frac{1-2n_{b'\!,\bi{p}'}}{2E_{b'\!,\bi{p}'}}\Delta_{b'\!,\bi{p}'},
\label{eq:gap}
\end{equation}
where
$E_{b,\bi{p}}=(\eta_{b,\bi{p}}^2+|\Delta_{b,\bi{p}}|^2)^{1/2}$ are
the quasiparticle energies and
$n_{b,\bi{p}}=[\exp(E_{b,\bi{p}}/k_{_{\mathrm{B}}}T)+1]^{-1}$ are
the Fermi filling factors with $k_{_{\mathrm{B}}}$ the Boltzmann
constant, and $T$ the temperature. The summation over the band
index $b'$ should be restricted to the partially filled (metallic)
bands, comprising sheets of the Fermi surface. Applying this
standard procedure to the HL Hamiltonian (\ref{eq:interaction}),
and after some algebra, we obtain the desired BCS pairing kernel,
\begin{eqnarray}
 V_{b,\bi{p};b'\!,\bi{p}'}
  =&-2\Jsd\, S_{\bi{p}} S_{\bi{p}'} D_{\bi{p}} D_{\bi{p}'}
  -J_{\mathrm{pp}}\,\gamma_x X_{\bi{p}} X_{\bi{p}'}\,
                      \gamma_y Y_{\bi{p}} Y_{\bi{p}'}\nonumber\\
  &+
 2\left(
      J_{\mathrm{sp}} S_{\bi{p}} S_{\bi{p}'} + J_{\mathrm{pd}} D_{\bi{p}} D_{\bi{p}'}
    \right)
    \left(
      \gamma_x X_{\bi{p}} X_{\bi{p}'}
    + \gamma_y Y_{\bi{p}} Y_{\bi{p}'}
    \right),
 \label{eq:Vpp}
\end{eqnarray}
where
\begin{equation}
 \gamma_x = 4\cos\left(\frac{p_x+p'_x}{2}\right),\qquad
 \gamma_y = 4\cos\left(\frac{p_y+p'_y}{2}\right).
\label{eq:gxgy}
\end{equation}
As the band indices $b$ and $b'$ enter implicitly in the band
energies $\eps_{b,\bi{p}}$ in the equation for the eigenvectors
$\Psi_{\bi{p}}(\eps_{b,\bi{p}})$, we will suppress them hereafter.
The layered cuprates, admittedly, have a single conduction band
and their Fermi surface has the shape of a rounded square. In this
simplest case one has to solve numerically the nonlinear integral
equation
\begin{equation}
 \Delta_{\bi{p}} = \int\limits_{-\pi}^{\pi}\frac{\rmd q_x}{2\pi}
 \int\limits_{-\pi}^{\pi}\frac{\rmd q_y}{2\pi}\,(-V_{\bi{p}\bi{q}})
 \frac{\Delta_{\bi{q}}}{2E_{\bi{q}}}\,
  \tanh\!\left(\frac{E_{\bi{q}}}{2k_{_{\rm B}}T}\right).
\label{eq:gapa}
\end{equation}
The solution to this general gap equation, depending on the
$J_{ij}$ values, can exhibit s, p, or d-type symmetry. It has been
shown previously that a purely p-p
model~\cite{Mishonov:94,Mishonov:98} ($J_{\mathrm{pp}} > 0$)
results in a $\rmd_{xy}$ ($B_{2g}$) gap anisotropy.  However, we
found that an agreement with the experimentally observed
$\rmd_{x^2-y^2}$ ($B_{1g}$) gap anisotropy (for a review see for
example \Rref{TsueiKirtley}) can be achieved only in the simplest
possible case of a dominant s-d exchange. This separable
Hamiltonian deserves special attention and we will analyze it in
the next sections.

\section{Separable s-d model}
\label{sec:sd}

For the special case of a purely s-d model,
$J_{\mathrm{sp}}=J_{\mathrm{pd}}=J_{\mathrm{pp}}=0,$ representing
the spin exchange operator $\hat P$ as a $(4\times 4)$-matrix, cf.
equations~(\ref{eq:P})--(\ref{eq:bond}), the reduced pairing
Hamiltonian takes the form
\begin{equation}
\fl \hat{\cal{H}}_{\mbox{\scriptsize HL-R}}= \frac{\Jsd}{N}
\sum_{\bi{p},\bi{q}} \left(\!\!
    \begin{array}{c}
       \hat{S}_{-\bi{p}  \uparrow} \hat{D}_{\bi{p}  \uparrow}\\
       \hat{S}_{-\bi{p}  \uparrow} \hat{D}_{\bi{p}\downarrow}\\
       \hat{S}_{-\bi{p}\downarrow} \hat{D}_{\bi{p}  \uparrow}\\
       \hat{S}_{-\bi{p}\downarrow} \hat{D}_{\bi{p}\downarrow}
    \end{array}
 \!\!\right)^{\!\!\dag}
\left(
        \begin{array}{cccc}
           1 & 0 & 0 & 0 \\
           0 & 0 & 1 & 0 \\
           0 & 1 & 0 & 0 \\
           0 & 0 & 0 & 1
        \end{array}
    \right)
\left(\!\!
    \begin{array}{c}
       \hat{S}_{-\bi{q}  \uparrow} \hat{D}_{\bi{q}  \uparrow}\\
       \hat{S}_{-\bi{q}  \uparrow} \hat{D}_{\bi{q}\downarrow}\\
       \hat{S}_{-\bi{q}\downarrow} \hat{D}_{\bi{q}  \uparrow}\\
       \hat{S}_{-\bi{q}\downarrow} \hat{D}_{\bi{q}\downarrow}
    \end{array}
 \!\!\right) .
\end{equation}
Carrying out an additional reduction for a spin-singlet pairing,
the interaction Hamiltonian reads
\begin{equation}
 \hat{\cal{H}}_{\mbox{\scriptsize HL-R}}=
 -\frac{\Jsd}{N}\sum_{\bi{p},\bi{q},\alpha}
   \hat{S}_{\bi{p},\alpha}^\dag
   \hat{D}_{-\bi{p},-\alpha}^\dag
   \hat{D}_{-\bi{q},-\alpha}
   \hat{S}_{\bi{q},\alpha},
\end{equation}
where $-\alpha$ stands for the electron spin projection opposite
to $\alpha$. For comparison, we provide again the kinetic energy
part of the Hamiltonian employing the same notation,
\begin{equation}
\fl \hat{\cal{H}}_{\rm BH}'= \sum_{\bi{p},\alpha} \left(\!\!
    \begin{array}{c}
        \hat{D}_{\bi{p},\alpha} \\
        \hat{S}_{\bi{p},\alpha} \\
        \hat{X}_{\bi{p},\alpha} \\
        \hat{Y}_{\bi{p},\alpha}
    \end{array}
 \!\!\right)^{\!\!\dag}
\left(
        \begin{array}{cccc}
           \eps_{\mathrm{d}}-\mu & 0         & \tpd s_x      & -\tpd s_y     \\
                 0    &\eps_{\mathrm{s}}-\mu & \tsp s_x      & \tsp s_y      \\
             \tpd s_x & \tsp s_x  & \eps_{\mathrm{p}} - \mu  & -\tpp s_x s_y \\
            -\tpd s_y & \tsp s_y  & -\tpp s_x s_y & \eps_{\mathrm{p}} - \mu
        \end{array}
    \right)
\left(\!\!
    \begin{array}{c}
        \hat{D}_{\bi{p},\alpha} \\
        \hat{S}_{\bi{p},\alpha} \\
        \hat{X}_{\bi{p},\alpha} \\
        \hat{Y}_{\bi{p},\alpha}
    \end{array}
 \!\!\right)\! .
\end{equation}

\begin{figure}[t]
\centering\includegraphics[width=0.7\columnwidth]{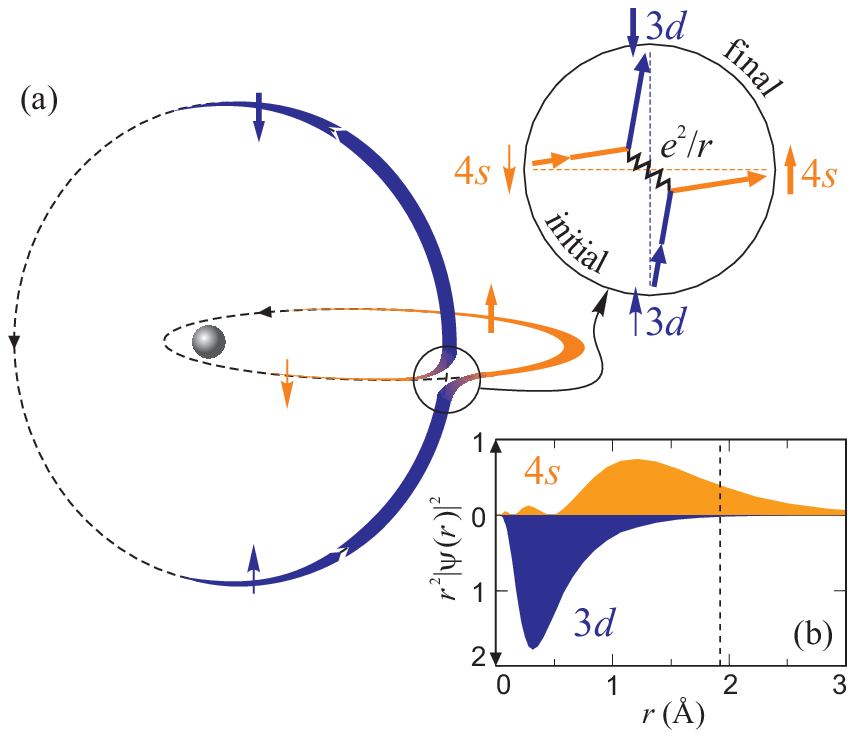}
\caption{Pairing two-electron exchange amplitude \Jsd\ ``hidden'' in the Cu
atom.  (a) Classical Bohr-Sommerfeld representation of the s-d two-electron
exchange process. The inset shows how the Coulomb scattering leads to an
effective electron spin exchange.  (b) Electron charge distribution for Cu~4s
and Cu~3d orbitals: the dashed line marks the Cu-O distance in the \cuo\
plane.
\label{fig:2} }
\end{figure}

Within the s-d model considered, the pairing kernel (\ref{eq:Vpp})
factors into functions depending only on $\bi{p}$ or $\bi{q},$
\begin{equation}
 (-V_{\bi{p}\bi{q}})
 = 2\Jsd\, S_{\bi{p}} D_{\bi{p}}\, S_{\bi{q}} D_{\bi{q}}
 \equiv 2\Jsd\, \chi_{\bi{p}}\,\chi_{\bi{q}}.
\label{eq:Vpq}
\end{equation}
A schematic representation of the \Jsd\ exchange amplitude is given in
\fref{fig:2}.  This factorizable Markowitz-Kadanoff~\cite{Markowitz:65} form
of the pairing kernel is a direct consequence of the local intra-atomic
character of the s-d exchange in the transition ion. Substituting in
equation~\eref{eq:gapa}
%
%
\begin{equation}
 \Delta_{\bi{p}}(T)=\Xi(T)\,S_{\bi{p}} D_{\bi{p}} = \Xi(T)\,\chi_{\bi{p}},
\label{eq:gapT}
\end{equation}
one obtains in a closed form, cf.~\Rref{Anderson:02}, a simple BCS equation
for the temperature dependence of the gap,
\begin{equation}
 \label{eq:gapsd}
 2\Jsd \left< \frac{\chi_{\bi{p}}^2}{2E_{\bi{p}}}\,
 \tanh\left(\frac{E_{\bi{p}}}{2k_{_{\rm B}}T}\right)\right>=1,
\end{equation}
where
\begin{eqnarray}
\label{eq:spectrum}
 E_{\bi{p}} & \equiv \left(\eta_{\bi{p}}^2+\Delta_{\bi{p}}^2\right)^{1/2}
  = \left[(\eps_{\bi{p}}-E_{\mathrm{F}})^2
              +(\Xi(T)\,\chi_{\bi{p}})^2 \right]^{1/2},\\
 \left< f_p \right> & =
 \int_0^{2\pi}\!\!\int_0^{2\pi}\frac{\rmd p_x\rmd p_y}{(2\pi)^2}\, f(\bi{p}),
\end{eqnarray}
$E_{\mathrm{F}}\equiv\mu.$ We wish to mention that separability of the order
parameter \eref{eq:gapT} has been derived by Pokrovsky~\cite{Pokrovsky:61} in
the general weak-coupling case and not only for factorizable pairing kernels.

According to \eref{eq:vector} we have
 \begin{eqnarray}
 \label{gap_exact}
 \fl \chi_{\bi{p}} \equiv S_{\bi{p}} D_{\bi{p}} =
   4\varepsilon_p\,\tsp\,\tpd\,(x-y) \left[\varepsilon_s\,\varepsilon_p^2-4
   \varepsilon_p\,\tsp^2\,(x+y)+32 \tpp\,\tau^2_{sp}\, xy\right]\nn\\
 \fl\qquad \times\left\{
   \left[4\varepsilon_p\,\tsp\,\tpd\,(x-y)\right]^2
  +\left[\varepsilon_s\,\varepsilon_p^2-4
   \varepsilon_p\,\tsp^2\,(x+y) + 32 \tpp\,\tau^2_{sp}\, xy\right]^2\right.\nn\\
 \fl\qquad\quad\left. +4x\left[(\varepsilon_s\,\varepsilon_p
   - 8\tau^2_{sp}\,y)\,\tpd\right]^2
 +4y\left[(\varepsilon_s\,\varepsilon_p -  8\tau^2_{sp}\,x)\,\tpd\right]^2
   \right\}^{-1}.
 \end{eqnarray}
The gap symmetry is then easily made obvious in the narrow-band
approximation. Formally, it is the asymptotic behaviour of the
eigenvector (\ref{eq:vector}) for vanishing hopping integrals
$t\rightarrow 0$. In this limit case~\cite{Mishonov:00}, we have
$\epsilon_{3,\bi{p}}\approx \eps_{\mathrm{d}},$ and
\begin{equation}
\tilde\Psi_{3,\bi{p}}= \left(\!\!
    \begin{array}{c}
        D_{3,\bi{p}} \\
        S_{3,\bi{p}} \\
        X_{3,\bi{p}} \\
        Y_{3,\bi{p}}
    \end{array}
\!\!\right) \approx \left(\!\!
    \begin{array}{c}
         1\\
        -(\tsp\tpd/\es\ep)\,(s_x^2-s_y^2) \\
         (\tpd/\ep)\, s_x \\
         (\tpd/\ep)\, s_y
    \end{array}
\!\!\right). \label{eq:vector3}
\end{equation}
%
\begin{figure}[t]
\centering
\includegraphics[width=0.45\columnwidth]{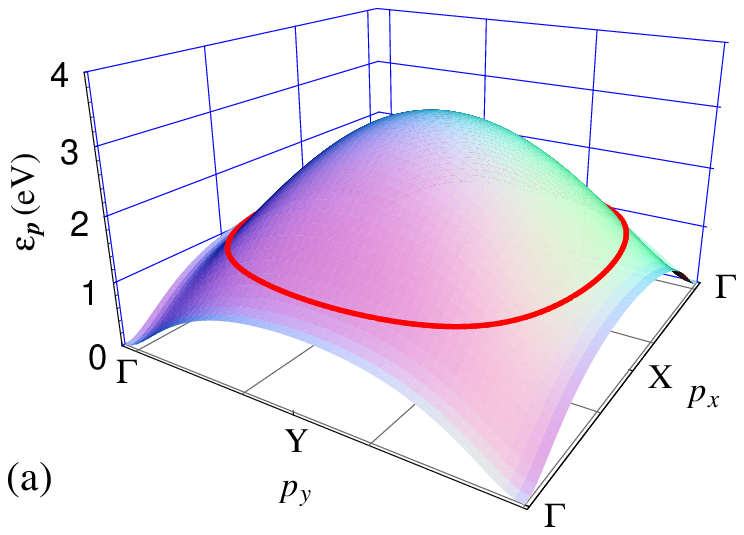}\hfill
\includegraphics[width=0.45\columnwidth]{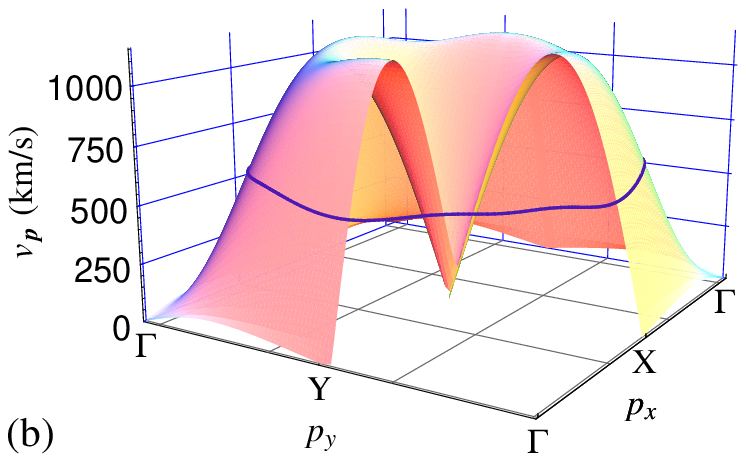}\\
\includegraphics[width=0.45\columnwidth]{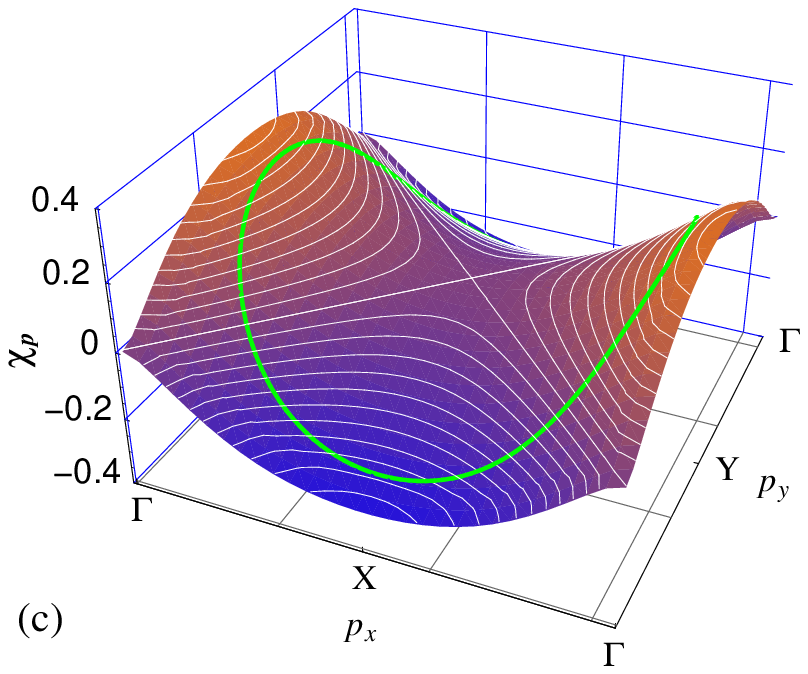}\hfill
\includegraphics[width=0.45\columnwidth]{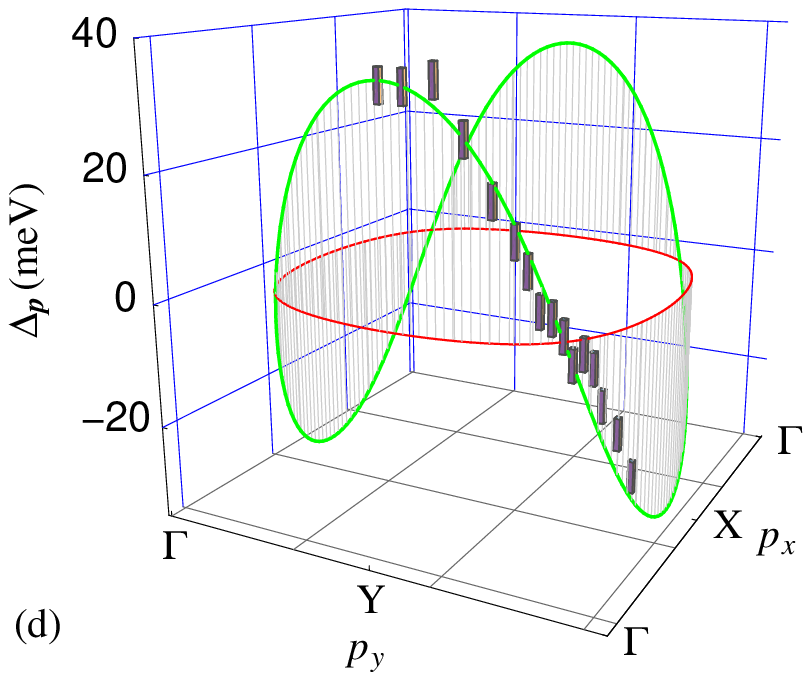}
\caption{Electronic properties of the superconducting \cuo\ plane.  (a)
Conduction band energy $\eps_{\bi{p}}$ as a function of the quasi-momentum
$\bi{p}.$ The red contour corresponding to the Fermi energy, $\eps_{\bi{p}} =
E_{\mathrm{F}},$ is in excellent agreement with the ARPES
data~\protect\cite{Schwaller:95}. (b) Quasiparticle velocity $v_{\bi{p}}$ as a
function of quasimomentum. The velocity variation along the Fermi contour is less
than 10~\%. The energy parameters are fitted to be in agreement
with the typical \textit{ab initio} calculations~\cite{bilayer-theor}. The
significant overestimate disappears if the bandwidth is fitted to the
experimental data, but the shape is conserved. (c) Momentum dependence of the
gap-anisotropy function $\chi_{\bi{p}}$ within the s-d model. The functional
values along the Fermi contour are indicated by a green line. (d)
Superconducting gap at zero temperature $\Delta_{\bi{p}}$ (green line)
according to our analytical result~\protect(\ref{eq:gapT}), plotted along the
Fermi contour (red line). The ARPES data~\protect\cite{Mesot:99} for
Bi$_2$Sr$_2$CaCu$_2$O$_{8+\delta}$ are given as prisms with sizes
corresponding to the experimental error bars. The gap function along the Fermi
contour has the same qualitative behavior and symmetry as the
Cu~3d$_{x^2-y^2}$ electron wave function along the circular orbit sketched in
\protect\fref{fig:2}(a). \label{fig:3}}
\end{figure}
%
Clearly, $D_{3,\bi{p}}$ exhibits $A_{1g}$ symmetry, while
$S_{3,\bi{p}}$ has the $B_{1g}$ symmetry.  Whence the product
$S_{3,\bi{p}}D_{3,\bi{p}}\propto \cos p_x-\cos p_y$ ``inherits''
the $B_{1g}$ symmetry, \fref{fig:3}(b), which is conserved even
for realistic values of the hopping integrals, and from
(\ref{eq:gapT}) it follows
\begin{equation}
 \label{gap_interpolation}
 \Delta_\bi{p}\propto S_{3,\bi{p}}D_{3,\bi{p}} \approx
 \frac{2\tsp\tpd}{\left(E_{\mathrm{F}}
 -\eps_{\mathrm{s}}\right)\left(E_{\mathrm{F}}-\eps_{\mathrm{p}}\right)}
 (\cos p_x -\cos p_y) .
\end{equation}
As can be seen in \fref{fig:3}(c) this small-$t$ approximation
fits the ARPES data for the gap anisotropy quite well. Similar
experimental data have been previously reported, e.g., in
\Rref{gapARPES}. Note, additionally, that close to the
$(\pi,\pi)$-point, where $(p_x-\pi)^2+(p_y-\pi)^2 \ll 1,$ the
angular dependence of the gap can be written in the form
\begin{equation}
 \label{eq:d-type}
 \fl
 \Delta_\bi{p}\propto \cos p_x -\cos p_y\approx
 \left[(p_x-\pi)^2+(p_y-\pi)^2\right] \cos 2 \phi,\qquad \tan \phi =
 \frac{p_y-\pi}{p_x-\pi}.
\end{equation}

The d-type angular dependence of both the gap anisotropy and the
separable pairing kernel is often postulated in phenomenological
model Hamiltonians to describe high-\tc\ superconductivity. The
previous discussion thus provides a microscopic justification
based on the fundamental exchange amplitudes.
For the oxygen scenario~\cite{Mishonov:00,Abrikosov:90}, in which
the Fermi level falls in a nonbonding oxygen band,
$\epsilon_{2,\bi{p}}\approx \epsilon_{\mathrm{p}}$ and
$t\rightarrow 0$~\cite{Mishonov:97}, the gap function has
different or additional nodes along the Fermi contour,
\begin{equation} \fl
 \left(\!\!
    \begin{array}{c}
        D_{2,\bi{p}} \\
        S_{2,\bi{p}} \\
        X_{2,\bi{p}} \\
        Y_{2,\bi{p}}
    \end{array}
 \!\!\right) \approx
   \frac{1}{\sqrt{s_x^2+s_y^2}}\left(\!\!
    \begin{array}{c}
     -2(\tpd/\ed)\,s_x s_y\\
      2\tsp(\tpp\ed+2\tpd^2)(\ed\tsp^2+\es\tpd^2)^{-1}
      \,s_x s_y(s_x^2-s_y^2) \\
     -s_y\\
      s_x
    \end{array}
 \!\!\right).
\label{eq:vector3b}
\end{equation}
Here $D_{2,\bi{p}}$ and $S_{2,\bi{p}}$ exhibit $B_{2g}$ and $A_{2g}$
symmetries, respectively. Let us also mention that the $\bi{s}$-vector
components, equation~\eref{eq:s}, constitute the arguments of the basis
functions of the symmetry representations.

Employing the analytical expression (\ref{eq:cec}) for the constant-energy
contours (CEC), one can implement an efficient numerical integration,
\begin{equation}
 \int\limits_0^{2\pi}\!\!\rmd p_x\!\int\limits_0^{2\pi}\!\!\rmd p_y\,
                                                     f(\eps_{\bi{p}})
 =\int\limits_{\eps_\mathrm{b}}^{\eps_\mathrm{t}}\!\!\rmd\eps
 \oint\frac{\rmd p_l}{v_{\bi{p}}}\, f(\eps),
\end{equation}
where
\begin{equation}
\label{velocity}
 v_{\bi{p}} = \left|\frac{\partial \eps_{\bi{p}}}{\partial \bi{p}}\right|
 = \frac{
     \left[({\cal A}y+{\cal B})^2x(1-x)
               +({\cal A}x+{\cal B})^2y(1-y)\right]^{1/2}
   }{
     |{\cal A}'xy+{\cal B}'(x+y)+{\cal C}'|
   },
\end{equation}
with ${\cal A}'$, ${\cal B}'$ and ${\cal C}'$ being the energy derivatives of
the polynomials (\ref{eq:ABC}),
\begin{eqnarray}
 {\cal A}'(\eps)&= 16\left[2\tsp^2\tpp-2\tpd^2\tpp-\tpp^2(\ed+\es)\right],\nn\\
 {\cal B}'(\eps)&= -4(\tsp^2\ed+\tpd^2\es) -4\ep(\tsp^2+\tpd^2),\\
 {\cal C}'(\eps)&= \es\ep^2+\ed\ep^2+2\ed\es\ep. \nn
\label{eq:ABCprime}
\end{eqnarray}
Using these functions, the band spectrum, see \eref{eq:cec}, can be
obtained by Newton iterations
\begin{equation}
\eps_{\bi{p}}^{[i]} = \eps_{\bi{p}}^{[i-1]} - \frac{{\cal A} xy + {\cal B}
  (x+y)+{\cal C}}{{\cal A}' xy + {\cal B}' (x+y)+{\cal C}'}
\end{equation}
with initial approximation for the conduction band
$\eps_{3,\bi{p}}^{[0]}=\ed.$

The charge carrier velocity is $v_{\bi{p}}a_0/\hbar$, $a_0$ is the lattice
constant, $p_l$ the dimensionless momentum component along the CEC, and
$\eps_\mathrm{b}$ and $\eps_\mathrm{t}$ are the bottom and the top of the
conduction band, respectively, $\eps_\mathrm{b}\le\eps_{\bi{p}}\le
\eps_\mathrm{t}$. The canonic equation for the CEC \eref{eq:cec},
\begin{equation}
 {\cal A}\cos p_x\cos p_y -({\cal A}+2{\cal B})(\cos p_x+\cos
 p_y)+{\cal A}+4{\cal C}=0,
\end{equation}
can be cast in an explicit form
\begin{equation}
p_{y,1}(p_x)= 2\arcsin\sqrt{-\frac{{\cal B}x+{\cal C}}{{\cal
A}x+{\cal B}}},\quad p_{y,2}(p_x)=2\pi-p_{y,1}(p_x),
\label{eq:cecexpl}
\end{equation}
and for the length element $\rmd p_l$ we obtain
\begin{equation}
 \rmd p_l =
 \sqrt{1+\left(\frac{\rmd p_y}{\rmd p_x}\right)^2}\rmd p_x, \quad
 \left(\frac{\rmd p_y}{\rmd p_x}\right)^2
  = \frac{x(1-x)}{y(1-y)}\!
    \left(\frac{{\cal A}y+{\cal B}}{{\cal A}x+\cal{B}}\right)^2.
\end{equation}
The contour integration along the hole pocket
$\varepsilon_{\bi{p}}=\mathrm{const}$ centered at the $(\pi,\pi)$ point needs to be
performed only over one eight of the CEC
\begin{equation}
\oint \rmd p_l f(p_x,p_y)= 8\int_{p_d}^\pi f(p_x,p_y(p_x))
\left(\frac{\rmd p_d(p_x)}{\rmd p_x}\right)\rmd p_x,
\end{equation}
where
\begin{equation}
 x_d =\sin^2\left( \frac{p_d}{2} \right),\qquad
      {\cal A}x_d^2+2{\cal B}x_d+{\cal C}=0.
\end{equation}

\section{Antiferromagnetic character of \Jsd\ }
\label{sec:signJ}

Let us address now the atomic physics underlying the s-d pairing
mechanism. Within the framework of the Hartree-Fock (HF) theory the exchange
energy is given~\cite{Zeiger} as an integral of the \cus\ and \cud\ atomic
wave functions,
\begin{equation}
 -\Jsd^{\mathrm{(HF)}}=\int\!\!\int
 \psi_{\mathrm{s}}^*(\bi{r}_1)\psi_{\mathrm{d}}^*(\bi{r}_2)
 \frac{
   e^2
 }{
   \left|\bi{r}_1-\bi{r}_2\right|
 }
 \psi_{\mathrm{d}}(\bi{r}_1)\psi_{\mathrm{s}}(\bi{r}_2)\,
 \rmd\bi{r}_1 \rmd\bi{r}_2,
\end{equation}
and its sign corresponds to repulsion and depairing for singlet Cooper
pairs. Thus, one can formulate the following conceptual problem, emerging in
fundamental physics:
\begin{itemize}
 \item[(i)] is it possible, as in the case of a covalent bond, for
   two-electron correlations to trigger a change of the sign of the exchange
   amplitude?
 \item[(ii)] how can one adapt the Heitler-London idea to a transition ion
   perturbed by ligands?
\end{itemize}
There is no doubt that the solution to this problem will illuminate other
problems in the physics of magnetism as well. In brief, the enigma can be
stated as to whether the Heitler-London approximation for the exchange energy
may result in $\Jsd > 0,$ cf. \Rref{Zeiger}.  Let us recall that already in
1962 Herring~\cite{HerringFlicker} was advocating that ``antiferromagnetic
$J_{ij}$'s should be the rule, ferromagnetic $J_{ij}$'s the exception''.  For
the present, we can adopt the s-d model as a convenient microscopic
phenomenology of superconductivity in the \cuo\ plane. On the other hand, the
exchange amplitude \Jsd\ is an important ingredient in the physics of
magnetism as well.

Physics of magnetism certainly displays lots of subtleties, but for a
qualitative comparison let us trace the ``operation'' of the s-d exchange
amplitude \Jsd\ in the case of the simplest model for a ferromagnetic
metal. While for the \cuo\ plane the s-band is empty, for transition metals it
is the widest conduction band. The width of the d-band is significantly
smaller and thus, making a caricature of the ferromagnetic metals, we
completely neglect the width of the d-band. In this ``heavy Fermion"
approximation the d-electrons are considered as localized, and without
significant energy loss they can be completely spin polarized, $\langle \hat
n_{\mathrm{d}\uparrow}\rangle \approx 1$, $\langle
\hat{S}_{\mathrm{d},z}\rangle \approx \frac{1}{2}>0$. In this case the
self-consistent approximation applied to \eref{eq:mixed} gives
\begin{equation}
 2\hat{\bi{S}}_{\mathrm{d}}\cdot\hat{\bi{S}}_{\mathrm{s}}\approx
 \hat{S}_{\mathrm{d},z}\left(\hat n_{\mathrm{s}\uparrow}-\hat
 n_{\mathrm{s}\downarrow}\right) \approx \langle \hat{S}_{\mathrm{d},z}\rangle
 \left(n_{\mathrm{s}\uparrow}-n_{\mathrm{s}\downarrow}\right).
\end{equation}
Here $n_{\mathrm{s}\alpha}\equiv\langle\hat
n_{\mathrm{s}\alpha}\rangle$ denotes the average number of
s-electrons per atom with spin projection $\alpha.$ In order to
calculate these variables one has to take into account the
different filling of the s-bands with different polarizations, and
sum over the quasi-momenta.  Finally, the exchange energy per atom
reads
\begin{equation}
 E_{\mathrm{X}}= - \frac{1}{2}\Jsd
 (n_{\mathrm{s}\downarrow}-n_{\mathrm{s}\uparrow}) < 0.
\end{equation}
In the \cuo\ plane, positive values of the \Jsd\ parameter lead to singlet
superconductivity. For ferromagnetic metals, positive values of \Jsd\
correspond to polarization of the s-band opposite to d-state polarization,
$n_{\mathrm{s}\downarrow}-n_{\mathrm{s}\uparrow}>0.$ Thereby ferromagnetism
could be brought about by an exchange amplitude with a sign corresponding to
antiparallel spin polarization of s- and d-orbitals, cf. figure~4-15 of
reference~\cite{Feynman}. Thus the same sign of the s-d exchange amplitude
\Jsd\ can be at the origin of ferromagnetism, e.g., in Fe and Ni, and
superconductivity in the \cuo\ plane. This is perhaps the simplest scenario
for cuprate superconductivity based on the two-electron exchange processes.

According to a naive interpretation of Hund's rule the Kondo effect should not
exist. In the epoch-making paper~\cite{Kondo} on the resistance minimum in
dilute magnetic alloys Kondo concluded that in the s-d exchange model, due to
Zener~\cite{ref:s-d}, Kasuya and Yosida~\cite{RKKY}, the sign of the direct
exchange amplitude \Jsd\ must be antiferromagnetic. And vice versa, the
minimum disappears if \Jsd\ is ferromagnetic. Such minimum exists for many
magnetic metals and alloys and is another hint in favor of Herring's
argument~\cite{HerringFlicker} mentioned earlier.  In his analysis Kondo
speculates that \Jsd\ is a parameter whose sign and magnitude have to be
determined so as to fit the experiment, and concluded that antiferromagnetic
values of the order of eV are quite reasonable. For a review on the Kondo
problem we refer the reader to \Rref{Tsvelick:83}.

On the other hand every textbook on atomic physics tells us that parallel
electron spins and an antisymmetric wave function minimize the electrostatic
energy. Put differently, the tendency toward ferromagnetism in Hund's rule is
of electrostatic origin. As Kondo has pointed out~\cite{Kondo}, the problem is
to find the origin of an antiferromagnetic \Jsd, or how to overcome the strong
electrostatic repulsion. It is very plausible that it is not a single driving
force, but instead one has to take into account several interfering electron
scattering amplitudes.

\subsection{Intra-atomic correlations}

The self-consistent approximation has been known in celestial mechanics for
ages. Accordingly, the motion of a planet is averaged over its orbit. One has
to calculate then the potential created by this orbital-averaged motion and
perform a sum over all particles. Where does this scheme fail? It fails in the
case of a resonance when the periods for some planets are commensurate or just
equal. This is nothing but the case of a transition ion for which the energies
and classical periods are very close. Then the resonant repetitive electron
scattering, symbolically presented in~\fref{fig:2}, leads to strong electron
correlations like in the double Rydberg states of atoms~\cite{Read:82}. For
double Rydberg states in He it is necessary to solve a two-electron quantum
problem but for other atoms we have to take into account the influence of the
other electrons in some self-consistent approximation, the local density
approximation (LDA), for example. For two 4s electrons the two-electron
correlations are so strong that they have to be taken into account from the
very beginning\cite{Read:82}. There are no doubts that the two-electron
correlations between 4s and 3d electrons having almost equal energies cannot
be neglected. In other words Hartree-Fock theory cannot be used
directly. Hence, the Bohr picture is not merely a historical remark but rather
an indispensable ingredient of the contemporary physics of magnetism:
two-electron correlations can be important even in a \emph{single} atom. We
thus conclude that the two-electron correlations may overrule Hund's rule for
the local s-d exchange. Note also that the single-particle orbitals
(accessible, e.g., from DFT~\cite{DFT}, HF and $X_{\alpha}$ methods, etc.)
only form an adequate basis for a subsequent account of electronic
correlations. A first step in this direction will be \textit{ab initio}
calculation of the \Jsd.

\subsection{Indirect s-d exchange}

The antiferromagnetism of the insulating phase of the undoped cuprates is
mediated by the Bloch-Kramers-Anderson indirect exchange~\cite{Anderson:63}
between 3d electrons of nearest-neighbour Cu ions via O~2p electrons. It is unlikely
that the numerical value of this $J_{\mathrm{dd}}$ exchange integral be
dramatically changed in the metallic phase obtained by hole doping. In the
metallic phase, however, the same indirect exchange mechanism will operate
between 3d and 4s electrons at the same Cu atom via the 2p electrons of the O
ligands. For illustration, let us compare the indirect s-d exchange amplitude
$\Jsd^{\mathrm{(ind)}}$ with $J_{\mathrm{dd}}$. There are three important
factors: (i) Every Cu ion has four~O ligands, \fref{fig:1}~(a). (ii) The
hopping amplitude between 4s and 2p orbitals is bigger than the 3d-2p
transfer. (iii) The Cu on-site Coulomb repulsion between 4s and 3d electrons
$U_{\mathrm{sd}}$ is much smaller than the 3d-3d Hubbard repulsion
$U_{\mathrm{dd}}$. Taking into account these factors one can expect that
$\Jsd^{\mathrm{(ind)}}$ is an order of magnitude bigger than
$J_{\mathrm{dd}}$:
\begin{equation}
 \Jsd^{\mathrm{(ind)}} \simeq 4 \left(\frac{\tsp}{\tpd}\right)^2
 \frac{U_{\mathrm{dd}}}{U_{\mathrm{sd}}} J_{\mathrm{dd}}.
\label{eq:ind}
\end{equation}
The relatively small $J_{\mathrm{dd}}$ ensures N\'eel temperatures
$T_{\mathrm{N}}$ of the order of room temperature. Hence we can conclude that
the indirect exchange can contribute significantly to the total \Jsd\
amplitude responsible for the pairing. However, only very detailed
first-principles calculations can clarify the relative contributions of the
direct and the indirect s-d exchange.

\subsection{Effect of mixing wave functions}

In an early paper~\cite{Kondo:62}, by analyzing the $g$-shift and the
anomalous Hall effect in Gd metal, Kondo showed that an antiferromagnetic
\Jsd\ can result from the effect of mixing the wave functions of conduction
and d electrons. We believe that this property is preserved if the d electrons
also form a conduction band, or even in the case of a single s-p-d hybridized
band. We should note that Kondo's argumentation for the need of a \Jsd\ with
an antiferromagnetic sign in the Kondo effect is related to Anderson's
consideration of localized magnetic states in metals~\cite{Anderson:61}. In
the latter schematized model, based on the works of Friedel~\cite{Friedel},
Anderson shows that ``any $g$-shift caused by free-electron polarization will
tend to have antiferromagnetic sign.''

As it was expected by various investigators the later numerical calculations
confirmed that the striking features of negative hyperfine field with large
amplitude comes mainly from the contact contribution of the core
electrons~\cite{Muto:64,Freeman:66,Ganguly:73}. The antipolarization between
the s- and d-electrons in transition metal compounds is also well observed by
M\"ossbauer spectroscopy, however the contribution of the core s-electrons and
conduction band cannot be experimentally resolved. For the pairing, the
amplitude of the s-d Kondo scattering is essential because in some terminology
the \cuo\ plane can be considered to be a single-band Kondo lattice,
cf. \Rref{Anderson:02a}.

Given the above diversity of channels for s-d exchange it is not surprising
that an adequate first-principles scheme to calculate \Jsd\ is still
sought. Furthermore, \Jsd\ is involved in the theory of magnetism in an entangled
way precluding so far a direct relation between \textit{ab initio}
calculations and formulae fitting the experiment~\cite{Vollhardt}.

\subsection{Cooper and Kondo singlet formation}

Although Kondo~\cite{Kondo} does cite Zener's paper~\cite{ref:s-d}, in many
publications in the field of magnetism the s-d interaction is referred to as
Kondo interaction. Often this term is used in a broader sense causing
eventually terminological misunderstanding. Here we cite some works which
could be related to our theory even though the relation may not be direct or
immediately apparent. Analyzing the possible ``interplay of Cooper and Kondo
singlet formations in high-$T_c$ cuprates'' Sekitani \etal~\cite{Sekitani:01}
point out that ``In the 20$^{\textrm{th}}$ century, two significant many-body
phenomena due to spin singlet formation were discovered in the field of solid
state physics: superconductivity and the Kondo effect''. These authors believe
that the pseudogap in the normal state corresponds to the dissociation energy
of the Kondo bound states and that superconductivity and the Kondo effect are
competing in La$_{1.85}$Sr$_{0.15}$CuO$_4$. They speculate that the interplay
between superconductivity and the Kondo effect has not be considered seriously
for high-\tc\ cuprates (further references on the Kondo effect in cuprate
superconductors can be found, e.g., in~\cite{Park:03}).  It would be premature
for us to comment on these ideas; we note however that within this terminology
our theory could be considered as a Kondo interaction mechanism for pairing in
high-\tc\ cuprates.

It is unclear from microscopic point of view if the same ``Kondo interaction''
amplitude \Jsd\ is responsible for the empirical Kadowaki-Woods
ratio~\cite{Kadowaki:86} but the location of La$_{1.7}$Sr$_{0.3}$CuO$_4$ on
the $A$ (the coefficient of the $T^2$ resistivity) versus $\gamma_0$ (the
electronic specific-heat coefficient) plot, cf. figure~4 in~\cite{Nakamae:03},
is a serious hint that \Jsd\ in the \cuo\ plane is one of the largest exchange
amplitudes in solid state physics, comparable with the uranium heavy-fermion
compounds and SrVO$_3$. In this sense our theory requires a large, yet
acceptable \Jsd\ value, putting the cuprates among the most interesting
materials with considerable exchange interaction.

\section{Dogmatics, Discussion, Conclusions and Perspectives}

In a review on the history of studies of superconductivity and the
prospects for further research in the field
Ginzburg~\cite{Ginzburg:00} conditionally divided the history into
several periods:

\begin{itemize}
\item[(i)] The ``Day Before Yesterday'' (1911--1941).  This period
starts with the discovery of superconductivity in Leiden by
Gilles~Holst and
Heike~Kamerlingh~Onnes~\cite{Nobel:96,Ginzburg:00}.
\item[(ii)] ``Yesterday'' (1942--1986). This period embraces the
appearance of the $\Psi$-$\Delta$ theories and the first
significant technical applications.
\item[(iii)] ``Today'' (1987--?). This epoch emerged with the
discovery of the high-\tc\ cuprates~\cite{Bednorz:86}.
\item[(iv)] ``Tomorrow'' (?). The final landmark of ``Today'' must be
some event.
\end{itemize}

Long ago, in the ``Day Before Yesterday'' high-\tc\ superconductivity was
known as a ``\textit{blue dream}'' of physicists. Considerable theoretical
efforts were applied ``Yesterday'', attempting to predict possible
realizations of this phenomenon~\cite{Ginzburg-Yes}. At that time the problem
of high-\tc\ superconductivity was ``one of the most interesting and
attractive problems from the purely scientific point of
view''~\cite{Ginzburg-Yes}. Intriguingly, the special role of \emph{layered
metallic systems} and \emph{almost two-dimensional
superconductivity}~\cite{GinzburgKirzhnits} was mentioned already in this
epoch, and a big variety of mechanisms of superconductivity were considered
including the s-d exchange~\cite{Akhiezer:59,s-d_super}. This exchange process
was well known in the physics of magnetism since the dawn of quantum
mechanics. Thus it is not surprising that the first work on the s-d pairing
mechanism, by Akhiezer and Pomeranchuk~\cite{Akhiezer:59}, was accomplished
about a year after the celebrated BCS paper~\cite{BCS}. These pioneering
works, however, ``have been ignored thus far''~\cite{Matthias:64}.

After Bednorz and M\"uller's work~\cite{Bednorz:86} the problem of high-\tc\
superconductivity soon came into fashion.  ``\textit{After experiencing the
`smell of roast meat', yesterday skeptics or even critics can become zealous
advocates of a new direction of endeavor.  But this is another story---more in
the realm of psychology and sociology than scientific and technical
activity}''~\cite{Ginzburg:91}.  All models of high-\tc\ superconductivity
were revisited in great detail in the uncountable number of papers that have
appeared in the epoch ``Today''.

\subsection{Aesthetics and frustrations of the central dogmas}

The common trends of some new theoretical models for cuprate superconductivity
were systematized by Anderson~\cite{Anderson:94} in six dogmas. We find it
very instructive to compare our theory of high-\tc\ superconductivity with
these dogmas.

\textit{``Dogma I: All the relevant carriers of both spin and
electricity reside in the \cuo\ plane and derive from the
hybridized O~2p--\cud\ orbital which dominates the binding in
these compounds. \dots in summary look at the planes only (a great
and welcome simplification.)''}

The key ingredient of our pairing theory is the four-fermion s-d interaction
between the \cus\ and \cud\ orbitals. If we cut the \cus\ orbital off from the
Hilbert space of the \cuo\ plane such a pairing interaction cannot
exist. Although \cus\ is an empty band, it is an important component of the
theory of high-\tc\ superconductivity. The O~2p orbitals are the intermediaries
between the \cus\ and \cud\ orbitals that create the necessary s-d
hybridization of the conduction \cud\ band.

\textit{``Dogma II: Magnetism and high-\tc\ superconductivity are
closely related, in a very specific sense: i.e., the electrons
which exhibit magnetism are the same as the charge carriers. \dots
We must solve the old problem of doping of a single Mott--Hubbard
band before we can begin the problem of high-\tc.''}

The incommensurate spin-density waves (SDW) observed in the
superconducting phase of La$_2$CuO$_{4.11}$ and
La$_{1.88}$Sr$_{0.12}$O$_4$ by neutron scattering~\cite{neutron} and
muon spin relaxation~\cite{muSR}, respectively, demonstrated that
antiferromagnetism of the Cu site is innocuous for superconductivity
in the cuprates. These antiferromagnetic correlations are not
depairing and do not change significantly \tc\ and the electronic
structure of \cuo\ plane. The observed correspondence between the
magnetic and the superconducting order parameters is an additional
hint that both phenomena have a common origin~\cite{cohabitation}; see
also the detailed theoretical
works~\cite{Kirkpatrick:03}. Nevertheless the coexistence of SDW and
superconductivity with a common critical temperature cannot be clearly
observed in every high-\tc\ cuprate. As a result superconductivity can
be considered, at least in first approximation, separately from a
possible antiferromagnetism as is done in the present paper. In Cr
metal the amplitude of the SDW shows also a BCS-like temperature
dependence~\cite{Overhauser} and the SDW-theory is based on the
conventional theory of metals based in turn on the Landau Fermi-liquid
theory. We consider the quasiparticle picture as a reliable starting
point for the theory of high-\tc\ cuprates as well.

\textit{``Dogma III: The dominant interactions are repulsive and
their energy scales are all large. \dots Restrict your attention
to a single band, repulsive (not too big) $U$ Hubbard Model.''}

Indeed, the dominant interactions are repulsive---``Nobody has
abrogated the Coulomb law'', as Landau used to
emphasize~\cite{Ginzburg:00}. However, something subtle occurs when
the atomic orbitals are analyzed. The strong electron repulsion leads
to Heitler-London type correlations: two electrons cannot occupy
simultaneously the same orbital, not even if they have opposite spins.
The exchange of electrons between two orbitals decreases the electron
kinetic energy and thereby the total energy of the whole system. In
molecular physics, according to the Hellmann-Feynman theorem such a
decrease in energy drives an inter-atomic attraction for large
inter-atomic distances. Thus, the valence attraction is the final
result of the dominant Coulomb repulsion between electrons. In this
way the Heitler-London-type exchange between itinerant electrons gives
rise to electron-electron attraction and conventional Cooper
pairing. The s-d exchange, ``residing'' in the Cu atom, can be
considered as an "intra-atomic-valence bond"---an attraction-sign
scattering amplitude due to the Coulomb repulsion between the
correlated electrons. The s-d exchange in the transition ions is one
of the most intensive exchange processes in solid state physics.  Such
a high-frequency process is described by the exchange amplitude \Jsd\
in the lattice models for the electronic structure and its sign is
determined by the inter-electronic Coulomb repulsion. The
Heitler-London interaction is a result of strong electron repulsion
and survives even for infinite Hubbard $U.$ This interaction is lost
when starting with the infinite-$U$ Hubbard model, however.  Thus, not
a single-band Hubbard model but a single-band s-d model with
antiferromagnetic exchange amplitude is the adequate starting point
for a realistic treatment of \cuo\ superconductivity.

\textit{``Dogma IV: The `normal' metal above \tc\dots is not a
Fermi liquid\dots but retains a Fermi surface satisfying
Luttinger's theorem at least in the highest-\tc\ materials. We
call this a Luttinger Liquid.''}

Very recently, the crucial experiment has finally been conducted.
After 15 years of intensive investigations of the cuprates it is
now experimentally established~\cite{Wiedeman-Franz} that the
overdoped cuprates obey the 150-years-old Wiedemann-Franz law
within a remarkable 1\% accuracy. After this experimental
clarification the theoretical comprehension will hardly keep us
waiting long. This experiment has also solved the old problem of
the nature of charge carriers created by doping of a single
Mott-Hubbard band, cf.~\textit{Dogma~II}. Now we know that charge
carriers of the normal state are standard Landau
quasi-particles~\cite{Landau} for which we have conventional
Cooper pairing in the superconducting phase. ``Holons'',
``spinons'' and spin-charge separation are unlikely to occur and
behave so as to emulate the properties of the ideal Fermi gas.  As
a function of the hole doping per Cu atom, $\tilde p,$ the
critical temperature is a smooth parabola~\cite{Presland:91},
\begin{equation}
  \tc/\tc^\mathrm{max}\approx 1-82.6(\tilde p - 0.16)^2.
\label{eq:Tc-p}
\end{equation}
Thus, it is improbable that the nature of the carriers and pairing mechanism
can be dramatically changed in the optimal and underdoped regimes although a
number of new and interesting phenomena complicate the physics of the
underdoped cuprates.

In short, in our opinion the experimental validation of the
Wiedemann-Franz law in overdoped cuprates~\cite{Wiedeman-Franz} is a
triumph of the Landau~\cite{Landau} and Migdal concept of Fermi
quasi-particles (and Landau spirit of \emph{trivialism} in general)
and provides a refutation of the spin-charge separation in
cuprates~\cite{Anderson:00a}. Hence, the problem of deriving the
Wiedemann-Franz law for strongly correlated electrons in the \cuo\
plane has just been set in the agenda.  According to the Fermi liquid
theory~\cite{Pines:1966} interactions between the particles create an
effective self-consistent Hamiltonian. As Kadanoff~\cite{Kadanoff:00}
has pointed out, this idea was much developed by
Landau~\cite{Landau:41} and
Anderson~\cite{Anderson:63a}. Unfortunately, for high-\tc\ cuprates a
link is still missing between the Landau quasiparticle concept and the
one due to Slater that even scattering matrix elements can be
calculated from first principles.

\textit{``Dogma V: Nonetheless, enough directions have been probed
to indicate strongly that this odd-even splitting of \cuo\ planar
states doesn't exist. \dots The impact of Dogma~V, then, is that
the two-dimensional state has separation of charge and spin into
excitations which are meaningful only within their two-dimensional
substrate; to hop coherently as an electron to another plane is
not possible, since the electron is a composite object, not an
elementary excitation.''}

Within the single-particle approximation (section~\ref{sec:model}) the
bilayer band splitting is readily obtained from \eref{eq:cec} and
\eref{eq:cecexpl} by the replacements
\begin{equation}
 \eps_i \rightarrow \eps_i \pm t_{\perp,ii},\quad i=\mathrm{s,\, p,\, d},
 \label{eq:split}
\end{equation}
where $t_{\perp,ii}$ is the hopping amplitude between the $i$th orbitals in
the adjacent \cuo\ planes. In other words, the two constant energy curves due to the bilayer
splitting are described by the same equation~\eref{eq:cec}.  Since it is
plausible that $t_{\perp,\mathrm{ss}}$ dominates, from \eref{eq:vector3} one
finds
\begin{equation}
\Delta E_{\mathrm{bilayer}} \approx
 2t_{\perp,\mathrm{ss}}
 |S_{3,\bi{p}}|^2 \approx 22 \mbox{ meV } (\cos p_x -\cos p_y)^2,
\end{equation}
in agreement with references~\cite{bilayer,bilayer-theor}. The
numerical value of 22~meV has been reported for heavily overdoped
Bi$_2$Sr$_2$CaCu$_2$O$_{8+\delta}$ (BSCCO)~\cite{bilayer}. This
experiment, crucial for \textit{Dogma~V},
cf.~reference~\cite{smoking-guns-battery}, is another piece of
evidence in favor of the conventional behaviour of the electron
excitations in the (\cuo)$_2$ slab.  Since $\Delta
E_{\mathrm{bilayer}}$ is relatively small in comparison with the width
of the conduction band, it is another hint that even for bilayer
superconductors like BSCCO and YBa$_2$Cu$_3$O$_{7-\delta}$ (YBCO) the
analysis of a single \cuo\ plane is an acceptable initial
approximation. Similarly, for fitting the three-dimensional Fermi
surface of Tl$_2$Ba$_2$CuO$_{6+\delta}$ determined by angle
magnetoresistance oscillations~\cite{Hussey:03a} one can start with
the simplest possible tight-binding approximation
\begin{equation}
 \eps_i \rightarrow \eps_i + t_{\perp,ii} \cos p_z,\quad i
 =\mathrm{s,\, p,\, d}.
\label{eq:ei}
\end{equation}

\textit{``Dogma VI: Interlayer hopping together with the
``confinement'' of Dogma~V is either the mechanism of or at least a
major contributor to superconducting condensation energy.''}

The interlayer hopping which is understood as a single-electron process
definitely cannot be considered as a two-electron pairing interaction creating
the condensation energy. It is only one of the details when one concentrates
on the material-specific effects in high-\tc\ superconductors. The inter-slab
hopping between double (\cuo)$_2$ layers is a coherent Josephson tunnelling
responsible for the long-living plasma oscillations with frequency
$\omega_{\mathrm{pl}} < \Delta$. These plasma oscillations along with far
infrared transparency of the superconducting phase were theoretically
predicted~\cite{Mishonov:91} for BSCCO---one of the few predictions made for
high-\tc\ cuprates, cf. the postdiction \cite{Anderson:98}. After the
experimental observation~\cite{Tamasaku:92}, the plasma resonances associated
with the Cooper-pair motion soon turned into a broad research
field~\cite{Buisson:94a}. Subgap plasmons were predicted~\cite{Mishonov:90}
for conventional superconducting thin films as well, and shortly after
experimentally confirmed~\cite{Buisson:94b} for thin Al films on SrTiO$_3$
substrate. The relatively lagged development of the physics of this effect was
partially due to the false neglect of the longitudinal current response in the
classical works on microscopic theory.  Concluding, let us note that the
London penetration depth $\lambda$ can be considered as the Compton wave
length of the Higgs boson of mass $m_{\mathrm{H}}c^2 =
\hbar\omega_{\mathrm{pl}},$ but the overall contribution of the interlayer
hopping to the condensation energy is negligible.

\subsection{Discussion}

The band structure of the \cuo\ plane is now believed to be
understood. However, after 15 years of development a mismatch of a factor of
two or three between the \textit{ab initio} and the experimental spectroscopic
estimates for the single-electron hopping amplitudes $t,$ or the bandwidth,
tends to be interpreted rather as a state-of-the-art ``coincidence''. The
Heitler-London approach is well-known in quantum
chemistry~\cite{Pauling,Eyring}, and has been successfully used for a long
time in the physics of magnetism~\cite{Goodenough}. We hope that realistic
first-principles calculations aiming at the exchange integrals $J$ of the
\cuo\ plane can be easily carried out. Should they validate the correct
(antiferromagnetic) sign and the correct order of magnitude of \Jsd,\ we can
consider the theory of high-\tc\ superconductivity established. We stress that
the two-electron exchange, analyzed here, is completely different from the
double exchange considered in reference~\cite{deGennes}.

In order to compare the derived results with the experiment, it is necessary
that the tight-binding conduction band energy be fitted to the available ARPES
data. In doing so a few parameters have to be properly taken into account: the
Fermi energy $E_{\mathrm{F}},$ as determined from the total area of the hole
Fermi contour, the difference between the Fermi energy and the Van Hove
singularity, $E_{\mathrm{F}}-\eps(\pi,0),$ the difference between the Van Hove
singular point and the bottom of the conduction band at the $\Gamma$ point,
$\eps(\pi,0) -\eps(0,0)$. The fit may further allow taking into account a
possible realization of the Abrikosov-Falkovsky scenario,
cf. \Rref{Mishonov:00}. According to the latter, for $\ed<\ep<\es$ and
sufficiently small \tpd, the conduction band can be the narrow (nonbonding)
oxygen band having a perfect (within the framework of the four-band model)
extended Van Hove singularity. If the superconducting gap has a $B_{1g}$-type
symmetry, its maximum value along the Fermi contour, $\Delta_{\max}=\max
|\Delta_\bi{p}(T=0)|,$ determines the \Jsd\ exchange integral in the s-d
model. Thus, the temperature dependence of the gap, described by the function
$\Xi(T),$ and the overall thermodynamic behaviour and low frequency
electrodynamic response will be determined without free fitting parameters.

The fit to the extended Van Hove singularity as observed, e.g., in
reference~\cite{Gofron:94} also points to a relatively low-lying \cus\ level
and one needs to consider the minimum value of $\eps_{\mathrm{s}}$. Although
the 4s~band is completely empty (\cus\ level is above the Fermi level), a very
close location is not ``harmless'' and would necessarily lead to some
prediction for the optical behaviour. With some risk of opening the Pandora's
box, we should mention that the lowest position of the \cus\ level is
determined by the mid-infrared (MIR) response. According to this possible
interpretation, the broadly discussed maximum of the absorption in the MIR
range is due to 3d-4s interband transition: one electron in the conduction
band is excited by the light to the empty \cus\ band. It seems that, up to
now, there is no natural explanation of this MIR optical adsorbtion (for a
review see~\cite{Plakida:95}).

The derived gap anisotropy function~(\ref{gap_exact}) and its
interpolation~(\ref{gap_interpolation}) compared to the ARPES experiment
showed that the ``standard'' four-band model spanned on the \cud, \cus,
O~2p$_x,$ and O~2p$_y$ orbitals, with an antiferromagnetic s-d pairing
interaction, successfully describes the main features of the ARPES data: the
rounded-square-shaped Fermi surface, small energy dispersion along the
$(0,0)$-$(2\pi,0)$ line, and the d-type $(B_{1g})$ symmetry of the energy gap
$\Delta_{\bi{p}}$ along the Fermi contour.  According to the pairing scenario
proposed here, strong electron correlations ``drive'' the electron exchange
amplitudes. These inter- and intra-atomic processes occur on energy scales
unusually large for solid state physics.  However, the subsequent treatment of
the lattice Hamiltonian can be performed completely within the framework of
the traditional BCS theory. The criterion for applicability of the BCS scheme
is not given by the $J$ vs $t,$ but rather by the \tc\ vs
$E_F-\eps_\mathrm{b}$ relation. Taking into account the typical ARPES-derived
bandwidths, which are much bigger than \tc\, we come to the conclusion that
the BCS trial wave function~\cite{Bogoliubov} is applicable for the
description of superconductivity in the layered cuprates with an acceptable
accuracy if \tc\ does not significantly exceed room temperature.

It is worth adding also a few remarks on the normal properties of the layered
cuprates. Among all debated issues in the complex physics of the cuprates, the
most important one is perhaps that of the normal-phase kinetics. The
long-standing problem is whether the paring interaction dominates (or totally
determines) the mechanism of Ohmic resistance in the normal phase, as is the
case for conventional superconductors. Within the present theory this question
can be formulated as follows: does the s-d exchange interaction dominate the
scattering of the normal-state charge carriers above \tc? This is a solvable
kinetic problem whose rigorous treatment will be given elsewhere. Here we
shall restrain ourselves in providing only a qualitative discussion.

In electron-electron scattering, just like in traffic accidents, the crucial
effect comes from the backscattering in ``head-on'' collisions. For
backscattering (i.e., $\vartheta = \pi$) in the case of s-d interaction, it
turns out that the matrix elements entering the pairing amplitude are also
important. It can be easily realized that this amplitude vanishes along the
diagonals of the Brillouin zone $(0,0)$-$(\pi,\pi)$ (the $\Gamma$-M
direction). Thereby the cold spots on the Fermi contour correspond to the
zeros of $\Delta_{\bi{p}}$. And vice versa, the hot spots are associated with
a maximum gap along $(\pi,\pi)$--$(0,\pi)$ (the M-X direction). In this sense,
cuprates repeat the qualitative feature of the conventional superconductors,
with a maximal gap corresponding to maximal scattering on the Fermi surface.

All layered cuprates are strongly anisotropic and two-dimensional models give
a reasonable starting point to analyze the related electronic processes. Most
importantly, the picture of a layered metal brings in something qualitatively
new which does not exist for a bulk metal---the ``interstitial'' electric
field between the layers, like the one in any plane capacitor. The
thermodynamic fluctuations of this electric field and related fluctuations of
the electric potential and charge density constitute an intensive scattering
mechanism analogous to the blue-sky mechanism of light scattering by density
fluctuations. It has recently been demonstrated~\cite{BlueSky} that the
experimentally observed linear resistance can be rationalized in terms of the
plane capacitor scenario; density fluctuations in the layered conductors are
more important than the nature of the interaction. In such a way the linear
normal-state resistivity is an intrinsic property~\cite{BlueSky} of the
``layered'' electron gas and cannot be used as an argument in favor of
non-Fermi-liquid behavior. The resistance of the normal phase may not be
directly related to the pairing mechanism and these problems can be solved
separately.  Nevertheless it will be interesting to check whether the
anisotropic scattering in cuprates~\cite{Hussey:03,Kaminski:05,Yusof:02} can
be explained within the framework of the s-d pairing Hamiltonian.

The present theory can also predict a significant isotope effect in the
cuprates. Even though the \Jsd\ pairing amplitude does not depend on the
atomic mass, the charge carriers reside the ionic \cuo\ 2D lattice, thereby
rendering polaron effects, as in any ionic crystal, possible. For the lighter
oxygen isotope the lattice polarization is more pronounced, leading to
enhanced effective mass and density of states, and reducing the transfer
integrals.  Overall, the isotope effect in the \cuo\ plane is due to the
isotope effect of the density of states. This rationale can be quantitatively
substantiated.  In our theory, upon isotope substitution at $T=0$, e.g., the
change of the penetration depth would be mainly driven by the Cooper pair
effective mass that could be determined by means of the Bernoulli effect. At
that the superfluid density remains unchanged. The calculation of the isotope
effect on \tc\ requires an evaluation of the polaron effects on the conduction
band.  Although this is a feasible problem, it is beyond the scope of the
present work.

The proposed mechanism for pairing in the \cuo\ plane can be handled much like
an ``Alice-in-Wonderland'' toy-model, but we find it fascinating that all
ingredients of our theory are achievements of quantum mechanics dating back to
the memorable 1920s, presently described in every physical textbook, and
constituting the fundamentals of solid state
physics~\cite{Ziman,Abrikosov}. It would be worthwhile attempting to apply the
approach, used in this paper, for modelling triplet and heavy-fermion
superconductivity as well.

\subsection{Conclusions: the reason for the success of the \cuo\ plane}

We find it very instructive to analyze qualitatively the reasons for the
success of the realization of high-\tc\ superconductivity in the \cuo\ plane:
\begin{itemize}
\item[(i)] Because of the relatively narrow quasi-two-dimensional conduction
d-band, due to p-d hybridization, the density of states is rather high. The
wide s-band resulting from s-p hybridization is completely empty, which is
somewhat unusual for compounds containing transition ions.
\item[(ii)] The pairing s-d exchange process was known since the first years
of quantum physics. It is omnipresent in the physics of the transition ions
but in order for it to become the pairing mechanism in perovskites it is
necessary that the s- and d-levels be close. In other words, a virtual
population of the s-level is at least needed in order to make the \Jsd\
amplitude operative.  Indeed, the conduction d-band is, actually, a result of
the s-p-d hybridization in the two-dimensional \cuo\ plane.
\end{itemize}

With the above remarks, one can speculate that among the perovskites the
layered ones are more favorable for achieving higher \tc (cf. also the
discussion in the Appendix on page~\pageref{app}).  The transition ion
series ends with Cu$^{2+}$ and the \cud\ and 4s levels are too close. One
should keep in mind that the filling of the electron shells finishes with a
``robbery'' in Cu~\cite{Feynman}: 3d$^{10}$4s$^{1}$ instead of 3d$^9$4s$^2$ as
one could expect from the electron configuration of the Ni atom 
(3d$^8$4s$^2$). However, the energy difference between these two Cu shell
configurations is very small.

Another favorable factor is the proximity of the O~2p and \cud\ levels.  Thus,
\textit{post factum} the success of Cu and O looks quite deterministic: the
\cuo\ plane is a tool to realize a narrow d-band with a strong s-p-d
hybridization. It was mentioned earlier that \Jsd\ is one of the largest
exchange amplitudes, but the 4s and 3d orbitals are orthogonal and necessarily
require an intermediary whose role is played by the O~2p orbital. Hence this
theory can be nicknamed ``\textit{the 3d-to-4s-by-2p highway to
superconductivity}''~\cite{highway}. The \Jsd\ amplitude is omnipresent for
all transition ion compounds, the hybridization of 3d, 4s and 2p is however
specific only for the \cuo\ plane.

How this qualitative picture can be employed to predict new superconducting
compounds is difficult to assess immediately. We believe, however, that this
picture, working well for the overdoped regime, is robust enough against the
inclusion of all the accessories inherent to the physics of optimally doped
and underdoped cuprates: cohabitation of superconductivity and
magnetism~\cite{cohabitation}, stripes~\cite{stripes},
pseudo-gap~\cite{pseudogap}, interplay of magnetism and superconductivity at
individual impurity atoms~\cite{Davis:01}, apex oxygen, \cuo\ plane dimpling,
doping in chains~\cite{Gonzalez:01}, the 41~meV resonance~\cite{Carbotte:99},
etc. Perhaps some of these ingredients can be used in the analysis of triplet
superconductivity in the copper-free layered perovskite
Sr$_2$RuO$_4$~\cite{Powell:02}. It is also likely that the superconductivity
of the RuO$_2$ plane is a manifestation of a ferromagnetic exchange integral
$J.$ The two-electron exchange mediates superconductivity and magnetism in
heavy Fermion compounds~\cite{heavyFermi} as well.  We suppose that lattice
models similar to the approach here will be of use in revealing the electronic
processes in these interesting materials.  Two-electron exchange may even
contribute to the 30~K \tc\ of the cubic perovskite Ba$_{0.6}$K$_{0.4}$BiO$_3$
but so far it is difficult to separate the exchange contribution from the
phonon part of the pairing interaction.  However, the strange doping behaviour
of Tl$_2$Ba$_2$CuO$_{6\pm\delta}$ in comparison with YBCO requires more
detailed investigation~\cite{Ruvalds:96}.

\subsection{Perspectives: if ``Tomorrow'' comes\dots}

The technological success in preparing the second generation of
high-\tc\ superconducting cables by depositing thin-layer
superconducting ceramics on a flexible low-cost metallic substrate is
crucial for the future energy applications. The USA Department of
Energy suggests global superconducting energy products would command
an annual market of 30~G\$ by about 2020. High-\tc\ superconductor
power cables, transformers, motors and generators could grab a 50\%
market share by 2013, 2015, 2016 and 2021,
respectively~\cite{Farkas}. On the other hand atomic-layer engineering
of superconducting oxides will trigger progress in materials science
and electronics. One can envision multi-functional all-oxide
electronics, e.g., sensors, processing and memory devices, all
monolithically integrated within a single chip~\cite{Bozovic}. In
spite of the technological progress and tens of thousands of
publications the theoretical ``\textit{picture in early 2000 remains
fairly cloudy on the whole}''~\cite{Ginzburg:00}. The landmark of
``Today'' must be some event. ``What event will it be? It is desirable
that this landmark be the insight into the mechanism of
superconductivity in high-\tc\ cuprates''~\cite{Ginzburg:00}.

In this paper we presented a traditional theory for
superconductivity in overdoped, and possibly also optimally doped
cuprates. All of its ingredients can be found in the textbooks and
there is a considerable chance that we witness the victory of
traditionalism, as it was in the history of quantum
electrodynamics (QED) half a century ago, but it may well be just
a personal viewpoint ``\textit{brainwashed by
Feynman}''~\cite{Anderson:00}.  Nonetheless let us use the example
of QED to illustrate the essence of our contribution. QED appeared
as a synthesis between perturbation theory and relativity. Both
components had been known well before the QED conception.
Similarly, both the BCS theory and the exchange interaction have
been known for ages, so the point in the agenda was how to
conceive out of them the theory of high-\tc\ cuprates.  Such a
theory contains necessarily a big number of energy parameters
($E_{\mathrm{F}}$, $\eps_{\mathrm{s}}$, $\eps_{\mathrm{p}}$,
$\eps_{\mathrm{d}}$, \tsp, \tpp, \tpd, \Jsd, $J_{\mathrm{pd}}$,
$J_{\mathrm{sp}}$, $J_{\mathrm{pp}}$) which are difficult to
determine simultaneously\footnote{%
The gap-anisotropy fit in \fref{fig:3}(d) is quite robust
  against the choice of the parameters. To illustrate and emphasize the
  capability of the model we have used, for example, unrealistically big
  values of the hopping integrals: $\ed=0$, $\es=5$, $\ep=-0.9$, $\tpd=1.13$,
  $\tsp=1.63$, $\tpp=0.2$~eV. This set of parameters corresponds to band
  calculations but gives a factor 2--3 wider conduction band. If the band is
  fitted to the ARPES data \Jsd\ can be less than 1~eV. A realistic fit is
  deemed to be a subject of a collaboration with experimentalists.
} (for the current status
of the problem see for example \Rref{Rosner:99,Pajda:01}). The
first step will definitely be to use ARPES data in which the
spectrum is clearly seen and to neglect in a first approximation
the ``irrelevant'' inter-atomic exchange integrals
$J_{\mathrm{pd}}$, $J_{\mathrm{sp}}$ and $J_{\mathrm{pp}}$. In
this case, for a known normal spectrum one can determine \Jsd\
from \tc\ or from the maximum gap at $T=0.$

A crucial ``meeting point'' between theory and experiment is the
Ginzburg-Landau (GL) theory. The general form of the GL coefficients for
anisotropic-gap superconductors, including the effect of disorder, is given in
\Rref{MPIP:02} and is directly applicable to the present model.  For the s-d
separable kernel~\eref{eq:Vpq} the specific heat $C(T)$ in the clean limit can
also be explicitly derived~\cite{C-T} and has the GL form
$C(T)=C_{\mathrm{N}}+C_\Delta,$ with
\begin{eqnarray}
C_{\mathrm{N}}(T) &= \frac{\pi^2}{3} \langle q_c(\nu_{\bi{p}})\rangle,\nn\\
C_\Delta (T)&= k_{\mathrm{B}}T \frac{\alpha^2}{b}
   = \frac{4\pi^2}{7\zeta(3)}
   \frac{\langle\chi_{\bi{p}}^2\,
   q_a(\nu_{\bi{p}})\rangle^2}{\langle\chi_{\bi{p}}^4\,
   q_b(\nu_{\bi{p}})\rangle}\,\theta(\tc - T),
\end{eqnarray}
where $ \nu_{\bi{p}} = \frac{E_{\bi{p}}}{2\kb T},$ and
\begin{eqnarray}\fl
\alpha(T) = \frac{1}{2(\kb T)^2}\langle \chi^2_{\bi{p}}\,
q_a(\nu_{\bi{p}})\rangle, \qquad b(T) =
\frac{7\zeta(3)}{16\pi^2(\kb T)^3}\langle \chi^4_{\bi{p}}\,
q_b(\nu_{\bi{p}})\rangle,
\label{eq:GL} \\
\fl
 q_a (\nu) = \frac{1}{2\cosh^2\nu}, \quad
 q_b (\nu) = \frac{\pi^2}{14\zeta(3)}\frac{1}{\nu^2}
              \left(\frac{\tanh\nu}{\nu}-\frac{1}{\cosh^2\nu}\right), \quad
 q_c(\nu)  = \frac{6}{\pi^2}\frac{\nu^2}{\cosh^2\nu}.\nn
\end{eqnarray}
Accordingly, the jump of the specific heat at \tc\ is expressed by the GL
coefficients $\alpha$ and $b,$ $\Delta C = \kb\tc\alpha^2(\tc)/b(\tc).$ With
the help of the general equations \eref{eq:GL} one can further determine the
influence of the Van Hove singularity on the thermodynamic and electrodynamic
behaviour. For $\Delta C$ the effect of the Van Hove singularity is reported in
\Rref{ourVH}, and for a general review on the Van Hove scenario of high-\tc\
superconductivity we refer the reader to \Rref{VH}. When the Fermi level is
not close to the Van Hove singularity the GL coefficients can be worked out as
integrals over the Fermi surface; methodological details are given in
\Rref{Mishonov:02}.  Knowledge of the GL coefficients is also fundamental for
the physics of fluctuation phenomena in superconductors~\cite{Mishonov:00b}.

Furthermore, a microscopic consideration of the London penetration depth
$\lambda$ for screening currents in the \cuo\ plane gives~\cite{Edyn}
\begin{eqnarray}
  \frac{1}{\lambda^2(T)} &=
 \frac{e^2}{\varepsilon_0 c^2 \hbar^2 d_{\mathrm{eff}}} \oint v_{\bi{p}}
 r_d(\nu_{\bi{p}}) \frac{\rmd p_l}{(2\pi)^2}, \label{eq:lambda} \\
 r_d(\nu) &= \nu^2 \sum_{n=0}^{\infty} [\nu^2 + \pi^2(n + 1/2)^2]^{-3/2},\nn
\end{eqnarray}
where the integration is performed along the Fermi contour.
The penetration depth $\lambda(T)$ is involved in the
Bernoulli effect in superconductors~\cite{Bernoulli}:
\begin{equation}
  \frac{\Delta\varphi}{{\cal R}_{\mathrm{H}}} =
 - \frac{e^2}{2\varepsilon_0 c^2}\lambda^2(T)\, j^2, \qquad
  \frac{1}{{\cal R}_{\mathrm{H}}} = \frac{2|e|}{a_0^2 d_{\mathrm{eff}}}
  \oint p_y(p_x) \frac{\rmd p_x}{(2\pi)^2},
\label{eq:Rhall}
\end{equation}
where $\Delta\varphi$ is the change of the electric potential induced
by a current density $\mathbf{j},$ $1/{\cal R}_{\mathrm{H}}=e
n_\mathrm{tot}$ is the volume charge density of the charge carriers,
with $d_{\mathrm{eff}}$ the effective spacing between the \cuo\
planes.

For given penetration depth extrapolated to zero temperature,
$\lambda(0),$ and Hall constant of the superconducting phase, one can
easily determine the effective mass of the Cooper pairs
\begin{equation}\label{zinzm}
  m^* = \frac{e^*\lambda^2(0)}{\varepsilon_0 c^2{\cal R}_{\mathrm{H}}}, \qquad
  |e^*| = 2|e|.
\label{eq:m*}
\end{equation}
This important material parameter $m^*$ is experimentally accessible
from the electrostatic modulation of the kinetic inductance of thin
superconducting films~\cite{Mishonov:91b} as well as from the surface
Hall effect~\cite{Mishonov:99}.

Having a big variety of calculated variables the parameters of the
theory can be reliably fitted. Another research direction is the
first-principles calculation of the transfer amplitudes and
two-electron exchange integrals. The level of agreement with the
fitted values will be indicative for the completeness of our
understanding. In addressing more realistic problems, the properties
of a single space-homogeneous \cuo\ plane will be a reasonable
starting point. Concluding, we believe that there is a true
perspective for the theoretical physics of cuprate superconductors to
become an important ingredient of their materials science.

Magnetism and superconductivity are among the most important
collective phenomena in condensed matter physics. And, remarkably,
magnetism of transition metals and high-\tc\ superconductivity of
cuprates seem to be two faces of the same ubiquitous two-electron
exchange amplitude.

\ack

This work was supported by the Flemish GOA. This work is dedicated in
memoriam to our colleague and friend A~V~Groshev who was an
enthusiastic collaborator in the early stages of this years-long
endeavour. T~M~M is much indebted to T~Sariisky for the stimulating
discussions in the course of the pre-cuprates era seminars on the
problem of high-\tc\ superconductivity held in Sofia in the early
1980s. T~M~M is thankful to D~Damianov for partial financial support.
The realization of this work would have been impossible without the
cooperation of V~Mishonova. We are also indebted to Prof~M~Mateev for
his continuous support during this long research and to
Prof~B~Bioltchev for the support in the final stages of this
project. It is a pleasure to acknowledge extensive comments and
correspondence by (in chronological order) P~B~Littlewood, J~Zaanen,
C~Di~Castro, N~M~Plakida, J~M~J van~Leeuwen, J~de~Jongh,
P-G~de~Gennes, S~Sachdev, Ph~Nozi\`eres, C~M~Varma, V~L~Pokrovsky,
P~Wiegmann, B~L~Altschuler, A~Varlamov, A~Rigamonti, F~Borsa,
F~H~Read, M~Mateev, P~Brovetto, M~Sigrist, J~Bouvier, D~Damianov,
L~P~Pitaevskii, J~Friedel, J~Bok, M~Mishonov, D~Markowitz and
L~P~Kadanoff.


\appendix

\section*{Appendix. $\boldsymbol{\tc}$-$\boldsymbol{\eps_{\mathrm{s}}}$ correlations: a
  hint toward the mechanism of superconductivity in cuprates}
\addcontentsline{toc}{section}{Appendix. Tc-es correlations: a
  hint toward the mechanism of superconductivity in cuprates}
\label{app}

In the letter by Pavarini~\etal, \Rref{Pavarini:01}, a strong correlation is
observed between $T_{c\,\mathrm{max}}$ and a single parameter
\[
 s(\eps)=(\eps_{\mathrm{s}}-\eps)(\eps-\eps_{\mathrm{p}})/(2\tsp)^2,
\]
which is controlled by the energy of the Cu~4s orbital $\eps_{\mathrm{s}}$,
$\eps_{\mathrm{p}}$ is the O~2p single cite energy, \tsp\ is the transfer
integral between neighbor Cu $4s$ and O $2p$ orbitals, and $\eps_{\bi{p}}$ is
the conduction band energy whose bottom at the $\Gamma$ point coincides with
the Cu 3d$_{x^2-y^2}$ energy $\eps_{\mathrm{d}}$. It is unfortunate that
theorists have not so far paid any attention to this observation because it is
an important correlation between the \textit{ab initio} calculated parameter
$r=(1+s)/2$ and the experimentally measured $T_{\mathrm{c\ max}}$ which can
reveal the subtle link between the experiment and the theory and finally solve
the long-standing puzzle of the mechanism of high-\tc\ superconductivity
(HTSC).

The purpose of the present Comment~\cite{Mishonov04C} is to emphasize that the
missing link has already been found, and the work by Pavarini~\etal\ can be
used as a crucial test for theoretical models of HTSC.  Perhaps the simplest
possible interpretation, though one could search for alternatives, is given
within the framework of the present theory. In order for the
Schubin-Zener-Kondo exchange amplitude \Jsd\ to operate as a pairing
interaction of the charge carriers, it is necessary the Cu~4s orbital to be
significantly hybridized with the conduction band. The degree of this
hybridization depends strongly on the proximity of the Cu~4s level to the
Fermi level $\eps_{\mathrm{F}}$. Thus, it is not surprising that
$\eps_{\mathrm{s}}$ controls the maximal critical temperature $T_{\mathrm{c\
max}}$, being the only parameter of the CuO$_2$ plane which is essentially
changed for different cuprate superconductors.

Cu~3d and Cu~4s are orthogonal orbitals and their hybridization is
indirect. First, the Cu~3d$_{x^2-y^2}$ orbital hybridizes with the O~2p$_x$
and O~2p$_y$ orbitals, then the O~2p orbitals hybridize with Cu~4s. As a
result we have a ``3d-to-4s-by-2p'' hybridization of the conduction band of
HTSC cuprates which makes it possible the strong \emph{antiferromagnetic}
amplitude \Jsd\ to create pairing in a relatively narrow Cu~3d$_{x^2-y^2}$
conduction band. The hybridization ``filling'' of the Cu~4s orbitals can be
seen in cluster calculations as well~\cite{Stoll:03}. The $s(\eps)$ parameter
introduced in \Rref{Pavarini:01} reflects the proximity of all 3 levels in the
generic 4-band Hamiltonian of the CuO$_2$ plane---their ``random coincidence''
for the Cu-O combination. Suppose that those levels are not so close to each
other. In this case the slightly modified parameter
\begin{equation}
s'(\eps)=(\eps_{\mathrm{s}}-\eps)(\eps-\eps_{\mathrm{p}})/(4\tsp\tpd)
\end{equation}
is simply the energy denominator of the perturbation theory which describes
the hybridization filling of the axial orbital, see (4.10). Whence
$[s'(\eps)]^2$ is a denominator of the pairing amplitude in the BCS equation
(6.4). Hence, we conclude that the correlations reported in \Rref{Pavarini:01}
are simply the correlations between the critical temperature \tc\ and the
dimensionless BCS coupling constant $\rho(\eps_{\mathrm{F}})\Jsd / [s'(\eps_{\mathrm{F}})]^2$. Of
course, for coupling constants $\sim 1$ the BCS trial wavefunction can be used
only for qualitative estimates, but knowing the Hamiltonian the mathematical
problem may somehow be solved. In any case, even qualitatively, we are sure
that the stronger pairing amplitude \Jsd\ and hybridization $1/s'(\eps_{\mathrm{F}})$
enhance \tc.


Having LDA calculations for the band structures of many cuprates it is
worthwhile performing a LCAO fit to them~\cite{bilayer-theor} and using
experimental values of \tc\ to extract the pairing amplitude \Jsd\ for all
those compounds. The \textit{ab initio} calculation of the Kondo scattering
amplitude parameterized by \Jsd\ is an important problem which has to be set
in the agenda of computational solid state physics. We expect that it will be
a weakly material dependent parameter of the order of the s-d exchange
amplitude in Kondo alloys, but perhaps slightly bigger as for the Cu ion the
3d and 4s levels are closer compared to many other ions. Closer energy levels,
from classical point of view, imply closer classical periods of orbital motion
which leads to some ``resonance'' enhancement of the exchange amplitude due to
intra-atomic two-electron correlations. The final qualitative conclusion that
can be extracted from the correlations reported by Pavarini~\etal\ is the
explanation why only the CuO$_2$ plane renders HTSC possible, whereas hundreds
other similar compounds are not even superconducting, or have only a
``conventional'' value of \tc. The natural explanation is: because its
$s$-parameter is not small enough below its critical value. Even among the
cuprates one can find compounds with ``conventional'' values of \tc\ having
relatively large value of the $s$-parameter. For other transition metal
compounds the parameter
\[
 s'(\eps_{\mathrm{d}}) =
 (\eps_{\mathrm{s}}-\eps_{\mathrm{d}})(\eps_{\mathrm{d}}
 -\eps_{\mathrm{p}})/(4\tsp \tpd)
\]
is much bigger than its critical value $s_c$ which can be reached probably
only for the Cu-O combination. Thereby, the correlation reported by
Pavarini~\etal\ is a crucial hint which of the models for HTSC is still on the
arena. Concluding, we also note that the 4s hybridization is responsible for
the three-dimensional coherent Fermi surface of
Tl$_2$Ba$_2$CuO$_{6+\delta}$~\cite{Mishonov:05}

\end{document}